\documentclass[%
 reprint,
nofootinbib,
 amsmath,amssymb,
 aps,
]{revtex4-2}
\usepackage{xcolor}
\usepackage{mathtools}
\usepackage{braket}
\usepackage{soul}
\usepackage{amsmath}
\usepackage{textcomp}
\usepackage{relsize}
\usepackage{nicefrac}
\usepackage{graphicx}% Include figure files
\usepackage{dcolumn}% Align table columns on decimal point
\usepackage{bm}% bold math
\usepackage{hyperref}% add hypertext capabilities
\begin{document}

\preprint{APS/123-QED}

\title{Hermitian stochastic methodology for x-ray superfluorescence}% Force line breaks with \\

\author{Stasis Chuchurka}%
\email{stasis.chuchurka@desy.de}
\author{Vladislav Sukharnikov}%
\author{Nina Rohringer}
\email{nina.rohringer@desy.de}
\affiliation{
Deutsches Elektronen-Synchrotron DESY, 22603 Hamburg, Germany\;
\\
Department of Physics, Universität Hamburg, 22761 Hamburg, Germany
}%

\date{\today}% It is always \today, today,

% ηηηηηηηηηηηηηηηηηηηηηηηηηηηηηηηηηηηηηηηηηηηηηηηηηηηηηηηηηηηηηηηηηηηηηηηηηηηηηηηη

\begin{abstract}
    A recently introduced theoretical framework for modeling the dynamics of x-ray amplified spontaneous emission is based on stochastic sampling of the density matrix of quantum emitters and the radiation field, similarly to other phase-space sampling techniques. While based on first principles and providing valuable theoretical insights, the original stochastic differential equations exhibit divergences and numerical instabilities. Here, we resolve this issue by accounting the stochastic components perturbatively. The refined formalism accurately reproduces the properties of spontaneous emission and proves universally applicable for describing all stages of collective x-ray emission in paraxial geometry, including spontaneous emission, amplified spontaneous emission, and the non-linear regime. Through numerical examples, we analyze key features of superfluorescence in one-dimensional approximation. Importantly, single realizations of the underlying stochastic equations can be fully interpreted as individual experimental observations of superfluorescence.
\end{abstract}

% ηηηηηηηηηηηηηηηηηηηηηηηηηηηηηηηηηηηηηηηηηηηηηηηηηηηηηηηηηηηηηηηηηηηηηηηηηηηηηηηη

\maketitle

\section{Introduction}

Superfluorescence is observed when incoherently excited atoms collectively emit radiation in the form of a highly energetic, short burst of light. Since its theoretical discovery by Dicke \cite{1954'Dicke}, this phenomenon has been extensively explored in various experimental setups \cite{PhysRevLett.30.309, PhysRevLett.39.547, 1979'Vrehen,PhysRevLett.59.1189,Rain2018,Bradac2017,PhysRevLett.116.083601, PRXQuantum.3.010338}. Recently, it has gained renewed attention with the advent of x-ray free-electron lasers (XFELs) \cite{Rohringer2012, PhysRevLett.111.233902, Yoneda2015, PhysRevLett.120.133203, PhysRevLett.123.023201}. Focused XFEL beams prepare atoms in a state of sizable population inversion of inner-shell transitions by rapid inner-shell photoionization \cite{Rohringer2012, Yoneda2015, Duguay1967, Kapteyn1992, Kimberg2013}. The x-ray superfluorescence process starts from isotropic, spontaneous fluorescence {produced by incoherently excited atoms}, which, upon propagating through the excited medium, is exponentially amplified until saturation, resulting in short, directed x-ray emission bursts.

Subsequent to the initial realization of soft x-ray superfluorescence in Ne gas \cite{Rohringer2012, PhysRevLett.111.233902}, further experiments extended it to hard x-rays in solid targets \cite{Yoneda2015, 2022'Zhang}, and liquid jets \cite{PhysRevLett.120.133203, 2020'Kroll}. The emitted x-ray pulses have a small divergence, a pulse duration of a few femtoseconds, and improved coherence compared to the self-amplified spontaneous emission (SASE) XFEL pulses, and they exhibit high intensity, offering advantages in brilliance for x-ray spectroscopy. X-ray superfluorescence holds the potential as a source of x-ray radiation with unique characteristics. In Ref. \cite{2022'Zhang}, it has been demonstrated that double-pulse x-ray superfluorescence can be produced. Further refinement of this technique might pave the way for sources of phase-controlled x-ray pulses suitable for coherent nonlinear spectroscopy \cite{2017'Kowalewski}. Additionally, research in Ref.~\cite{2020'Halavanau} explores the possibility of creating high-brilliance x-ray laser oscillator sources.

Experimental studies and research into potential applications can greatly benefit from a quantitative theory of superfluorescence. A complete quantum description of superfluorescence is only feasible for small systems. { For the particular scenario of compact systems composed of identical multi-level emitters, we can adopt the idea behind symmetric Dicke states \cite{1954'Dicke}. Namely, we can decompose the quantum states into a basis set that maintains permutation symmetry. The number of resulting equations grows polynomially with the number of atoms \cite{gegg2016, 2018'Shammah, Sukharnikov2023, Silva2022}. Depending on the level structure of the emitters and involved incoherent processes, this enables treatment of hundreds of atoms. However, x-ray superfluorescence is typically observed from samples containing a macroscopic number of emitters. Additionally, proper consideration of propagation effects is necessary for x-ray superfluorescence. These effects break permutation symmetry, which is crucial for utilizing symmetric Dicke states. Thus, a different theoretical framework is required to address these aspects effectively.

Addressing the raised issues, Ref. \cite{Gross1982} extends Dicke's theoretical analysis to macroscopic extended samples.}  Specifically, under instantaneous excitation and the absence of incoherent processes, quantum effects can be replicated in Maxwell-Bloch equations supplemented by random initial conditions for the atomic variables. Their statistical properties are chosen so that averaging yields correct expectation values. However, if the initial incoherent excitation triggering superfluorescence is not instantaneous, and the quantum evolution is further complicated by various incoherent processes, random initial conditions are no longer applicable. In the x-ray range, the preparation of the population inversion, subsequent collective emission, and decoherence through Auger-Meitner decay happen on equivalent time scales, which requires other theoretical frameworks. In this case, several phenomenological strategies have been proposed. For instance, in Refs.~\cite{2004'Ziolkowski,2000'Larroche,2019'Subotnik-EhrenfestR,2019'Subotnik_comparison}, random initial conditions have been replaced by phenomenological noise terms acting as source terms of the Maxwell-Bloch equations. As these methods are not derived from first principles, they come with certain limitations. For example, the widely used methodology proposed in Ref.~\cite{2000'Larroche} produces an incorrect temporal profile of spontaneous emission, as highlighted in Refs.~\cite{2018'Krusic, 2019'Benediktovitch}.

In this article, we propose a versatile approach capable of describing all stages of x-ray superfluorescence in the presence of strong incoherent processes and pumping. Our method extends the formalism presented in Ref.~\cite{benediktovitch2023stochastic}, where the quantum master equation was transformed into a set of stochastic differential equations for the underlying dynamic variables. These equations are solved multiple times in a Monte Carlo manner, generating a sample of stochastic trajectories used to construct expectation values. The noise components of the stochastic equations represent the effect of quantum fluctuations or the stochastic nature of spontaneous emission.

A drawback of this stochastic approach, shared with other methods based on sampling the positive $P$ phase-space representation \cite{2014'Drummond_book, 2001CoPhC.142..442D, PhysRevE.96.013309, PhysRevLett.98.120402, PRXQuantum.2.010319}, is its tendency to produce diverging solutions \cite{2006'Deuar_stochastic-gauges, 2005'Deuar_PhD}. This issue arises because pairs of variables that are expected to be each other's complex conjugates in the classical limit do not maintain this property in the stochastic framework. In the context of superfluorescence, such dynamic variables include the positive and negative frequency components of the electric field or polarization density.

We examined this problem in the study of superfluorescence in compact systems \cite{Chuchurka2023compact}, where enforcing the Hermiticity of the underlying dynamic variables was proposed through the use of so-called stochastic gauges \cite{2006'Deuar_stochastic-gauges, 2005'Deuar_PhD}. These gauges allow for modifying the stochastic equations, albeit at the cost of properly re-weighting the stochastic trajectories. By comparing to the exact solution, this treatment proved very useful, nearly reproducing the exact solutions. However, a limitation is that the weighting coefficient involves an exponentiation by the number of particles in the system, potentially leading to instabilities for large systems. In such cases, satisfactory results can be achieved by simply skipping the re-weighting procedure (see Sec. V in Ref. \cite{Chuchurka2023compact}). This strategy was precisely the one adopted in the study of superfluorescence in extended systems \cite{benediktovitch2023stochastic}, though its effectiveness in the context of extended systems lacks direct evidence.

Any modification not rooted in a reasonable approximation runs the risk of distorting the statistical properties of superfluorescence. Here, we propose an alternative modification of the formalism, which relies on a reasonable approximation for extended, pencil-shaped media. This modification ensures the stability of the resulting equations while accurately describing the generation and amplification of spontaneous emission. 

{ Most of the spontaneous emission exits the elongated system immediately, and only a small fraction propagates along the sample. Although this fraction will be subsequently amplified and reach significant intensities, the generation of these initial spontaneous photons has a negligible impact on the atomic dynamics and can be treated perturbatively.  Additionally, during the initial phase of superfluorescence, it is crucial that incoherent processes exhibit a stronger influence on atomic dynamics than electromagnetic fields. Mathematically, the respective Rabi frequency is negligible compared to the inverse lifetime of the decaying states. This simplifies the expressions that govern the production of spontaneous emission, ensuring the Hermiticity of the underlying dynamic variables, thereby eliminating the issue of divergence. Upon propagation and amplification, electromagnetic fields intensify, and their effect on atomic variables becomes comparable to that of incoherent processes. However, at this stage, we assume that the fields are so strong that spontaneous emission can be completely neglected. }

In contrast to the formalism presented in Ref.~\cite{benediktovitch2023stochastic}, which yields complex-valued photon numbers at the level of individual stochastic trajectories, our approach eliminates such anomalies. This enables the interpretation of single realizations as individual observations from an experiment.

The provided framework is tailored for paraxial geometry and multi-level atoms. Yet, for a more focused examination of the statistical characteristics of superfluorescence, we demonstrate our approach via numerical simulations of x-ray superfluorescence in a one-dimensional geometry. As expected, the properties of spontaneous emission, which initiates superfluorescence, are fully reproduced. We furthermore extend our numerical study to also include an analysis of collective spontaneous emission in conjunction with a seed pulse.

Our developed formalism closely resembles the widely applied phenomenological approach of Larroche et al. \cite{2000'Larroche}, which serves as our benchmark for numerical comparison. {In contrast to this phenomenological approach, our derivations are grounded in first principles and reliable approximations, which enables us to accurately reproduce the temporal profile of spontaneous emission.} Notably, in the case of instantaneous excitation, our formalism reproduces the method based on random initial conditions \cite{Gross1982} or the concept of the tipping angle \cite{1979'Vrehen}.

The article is organized as follows: Section~\ref{sec: stochastic formalism} outlines the stochastic formalism derived in Ref.~\cite{benediktovitch2023stochastic}. Section~\ref{sec: linear} introduces key modifications to this formalism. Section~\ref{sec: self} formulates the resulting system of equations. Section~\ref{numerical examples} adopts these equations for the simplest system showcasing superfluorescence of a two-level system and provides numerical illustrations. In Sec.~\ref{finish}, we summarize the key findings and propose directions for future studies.

%%%%%%%%%%%%%%%%%%%%%%%%%%%%%%%%%%%%%%%%%%%%%%%%%%%%%%%%%%%%%%%%%%%%%%%%
\section{Stochastic methodology}\label{sec: stochastic formalism}
%%%%%%%%%%%%%%%%%%%%%%%%%%%%%%%%%%%%%%%%%%%%%%%%%%%%%%%%%%%%%%%%%%%%%%%%

 \begin{figure}[t!]
    \centering
    \includegraphics[width = 0.65\linewidth]{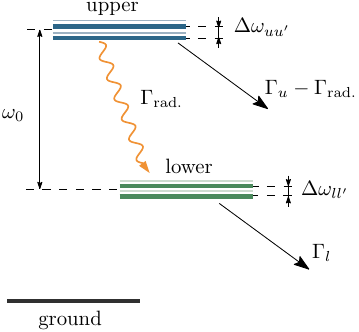}
    \caption{The stochastic formalism, described in Sec. \ref{sec: stochastic formalism}, is designed for atoms with the depicted level scheme. These levels are divided into three groups: ground, lower, and upper states. Initially, atoms reside in the ground state and are excited to the upper states through incoherent pumping. Subsequently, atoms transition from the upper states to the lower states, emitting spontaneous photons at the rate of $\Gamma_{\text{rad.}}$. The fraction traveling along the sample triggers superfluorescence. The energy difference between lower and upper states is denoted by $\hbar\omega_0$. Additionally, various incoherent processes affect the upper and lower states, resulting in inverse lifetimes $\Gamma_u$ and $\Gamma_l$.  }
    \label{fig: multi-level}
\end{figure}

In this section, we briefly revisit the stochastic formalism of Ref.~\cite{benediktovitch2023stochastic}, which provides detailed derivations of the formalism (Sec. III), implementation of pump-pulse propagation (Appendix A), and a list of decay rates and cross-sections for a copper sample (Appendix B).

The resulting system of equations resembles a set of Maxwell-Bloch equations modified with additional noise terms. These equations are formulated for a long and narrow gain medium in paraxial geometry typically encountered in x-ray superfluorescence experiments. Initially, the atoms are in their ground state. An x-ray pump pulse promotes atoms to core-excited states by photoionization of the inner electronic shell, achieving a population inversion in the produced ions. The excited states are represented by two manifolds as shown in Fig.~\ref{fig: multi-level}: upper states ${\ket{u}}$ and lower states ${\ket{l}}$ coupled by dipole moments $\textbf{d}_{ul}$ and $\textbf{d}_{lu}$. We decompose them into the product of the reduced dipole moment $d_0$~\cite{landau1977} and dimensionless coefficients $T_{ul,s}$ and $T_{lu,s}$ that are proportional to Clebsch-Gordan coefficients:
\begin{equation*}
\label{eq: d via T}
    \textbf{d}_{ul} \cdot {\textbf{e}}_s = d_0 T_{ul,s}, \quad
    \textbf{d}_{lu} \cdot {\textbf{e}}_s^* = d_0 T_{lu,s},
\end{equation*}
where ${\textbf{e}}_s$ represents the independent polarization vectors $s=1,2$ of the electromagnetic field. The upper $u$ and lower $l$ ionic state manifolds include levels with energy splittings,
\begin{align*}
    \Delta\omega_{uu'}&=\omega_u-\omega_{u'},
    \\
    \Delta\omega_{ll'}&=\omega_l-\omega_{l'},
\end{align*}
{that we assume are negligible in this study.} We reserve indices $u$ and $l$ for upper and lower states, respectively.

The ions, following photoionization of the neutral atoms, are characterized by the density matrix $\rho_{pq}(\textbf{r},\tau)$, with diagonal elements representing populations and off-diagonal elements representing coherences. In contrast to indices $u$ and $l$ reserved for upper and lower states, respectively, $p$ and $q$ represent any states. The paraxial geometry of the gain medium motivates the introduction of the retarded time $\tau = t - z/c$, conveniently incorporating propagation effects and facilitating numerical calculations. Additionally, we factorize out the fast oscillations $e^{\pm i\omega_0\tau}$ from the coherences $\rho_{ul}(\textbf{r},\tau)$ and $\rho_{lu}(\textbf{r},\tau)$ between upper and lower excited states.

When only electric dipole transitions are involved, the electromagnetic fields can be solely characterized by the electric field $\bm{\mathcal{E}}(\textbf{r},t)$, which we decompose into slowly varying amplitudes $\mathcal{E}_s^{(\pm)}(\textbf{r},t)$:
\begin{multline*}
   \bm{\mathcal{E}}(\textbf{r},t)=\sum_{s}\Big(\mathcal{E}_s^{(+)}(\textbf{r},t)\textbf{e}_s e^{-i\omega_0(t-z/c)}\\+\mathcal{E}_s^{(-)}(\textbf{r},t)\textbf{e}_s^* e^{i\omega_0(t-z/c)}\Big).
\end{multline*}
Furthermore, we introduce the Rabi frequencies
\begin{equation*}
    \Omega_s^{(\pm)}(\textbf{r},\tau)=\frac{d_0}{\hbar}\mathcal{E}_s^{(\pm)}(\textbf{r},\tau+z/c).
\end{equation*}
Here, we have used the retarded time $\tau$. More details on the definition of the dynamic variables, their connection to the quantum-mechanical operators, and the introduction of the retarded time can be found in Ref.~\cite{benediktovitch2023stochastic}.

The dynamics of the atomic variables $\rho_{pq}(\textbf{r},\tau)$ is defined by the modified Bloch equations, which we split into three parts. The first part describes the interaction with pump pulses and incoherent processes such as Auger-Meitner decay or isotropic fluorescence. Specifically, for the coherences, we write:
\begin{subequations}
\label{eq:atomic_equations}
    \begin{equation}
    \label{eq:atomic_equations_1}
    \frac{\partial}{\partial \tau}\rho_{p\neq q}(\textbf{r},\tau)\Big|_{\text{incoh.}}=
    -(\Gamma_{p}+\Gamma_{q})\rho_{pq}(\textbf{r},\tau)/2,
    \end{equation}
where $\Gamma_{p}$ and $\Gamma_{q}$ represent the inverse lifetimes of the states $\ket{p}$ and $\ket{q}$, respectively. In this article, we make the reasonable assumption that states originating from the same manifold share a common, stationary\footnote{Non-stationary lifetimes can be caused by, for example, secondary photoionization induced by the pump field.} lifetime (see Fig.~\ref{fig: multi-level}). The reason behind this assertion {as well as the assumption of negligible energy splittings $\Delta\omega_{uu'}$ and $\Delta\omega_{ll'}$ }will be clarified later in {Secs. \ref{sec: linear} and} \ref{sec: self}. Likewise, the first part of the equations for the populations are given by
\begin{multline}
\label{eq:atomic_equations_2}
   \frac{\partial}{\partial \tau}\rho_{pp}(\textbf{r},\tau)\Big|_{\text{incoh.}} =
    -\Gamma_{p}\rho_{pp}(\textbf{r},\tau)
    \\+p_{p}^{(\text{pump})}(\textbf{r},\tau)\rho^{(\mathrm{ground})}(\textbf{r},\tau)\\+\Gamma_{\text{rad.}}\sum_{k}G_{pk}^{(\text{rad.})}\rho_{kk}(\textbf{r},\tau),
\end{multline}
including the incoherent pump, populating the ionic states by photoionization of the atomic ground state $\rho^{(\mathrm{ground})}(\textbf{r},\tau)$. Here, $p_{i}^{(\text{pump})}(\textbf{r},t)$ represents the photoionization rate populating state $\ket{i}$. The third term describes population growth caused by radiative spontaneous decay. Its rate is proportional to $\Gamma_{\text{rad.}}=\omega_0^3 d_0^2/[3 \pi \varepsilon_0 \hbar c^3]$, and $G_{ik}^{(\text{rad.})}$ are dimensionless coefficients that depend on Clebsch-Gordan coefficients.

The second part of the generalized Bloch equations describes the interaction with the electromagnetic fields resonant with the excited states, reading:

\begin{widetext}
\begin{multline}
\label{eq:atomic_equations_3}
 \frac{\partial}{\partial \tau}\rho_{pq}(\textbf{r},\tau)\Big|_{\text{unitary}}=i\sum_{r,s}\Bigg[\Omega_s^{(+)}\!\left(\textbf{r},\tau\right)\Big(T_{p>r,s}{\rho}_{rq}\!\left(\textbf{r},\tau\right)-{\rho}_{pr}\!\left(\textbf{r},\tau\right)T_{r>q,s}\Big)\\+\Omega_s^{(-)}\!\left(\textbf{r},\tau\right)\Big(T_{p<r,s}{\rho}_{rq}\!\left(\textbf{r},\tau\right)-{\rho}_{pr}\!\left(\textbf{r},\tau\right)T_{r<q,s}\Big)\Bigg],
\end{multline}
{where indices $p$ and $q$ represent arbitrary upper or lower states. The summation over index $r$ encompasses all atomic states but is implicitly constrained by the conditions set on the subindices of the coefficients $T_{pr,s}$ and $T_{rq,s}$. For instance, $T_{p>r,s}$ indicates that the corresponding term is included in the sum only if $p$ corresponds to upper states ${\ket{u}}$ and $r$ corresponds to lower states ${\ket{l}}$. The same principle applies to $T_{r<p,s}$.} Finally, to account for spontaneous emission initiating superfluorescence, we introduce the following stochastic terms:
\begin{equation}
\label{eq:atomic_equations_4}
\begin{aligned}
    \frac{\partial}{\partial \tau}\rho_{pq}(\textbf{r},\tau)\Big|_{\text{noise}}&=\sum_s\bigg(\sum_{r}{\rho}_{pr}(\textbf{r},\tau)T_{r>q,s}-{\rho}_{pq}(\textbf{r},\tau)\sum_{u,l}T_{ul,s}{\rho}_{lu}(\textbf{r},\tau)\bigg)g_s^\dag(\textbf{r},\tau)\\&+\sum_s\bigg(\sum_{r}T_{p<r,s}{\rho}_{rq}(\textbf{r},\tau)-{\rho}_{pq}(\textbf{r},\tau)\sum_{u,l}T_{lu,s}{\rho}_{ul}(\textbf{r},\tau)\bigg) f_s^\dag(\textbf{r},\tau),
\end{aligned}
\end{equation} 
\end{widetext}
\end{subequations}
which involve independent noise terms \(f_s^\dag(\textbf{r},\tau)\) and \(g_s^\dag(\textbf{r},\tau)\) {whose statistical properties are discussed later}. Similarly, noise terms are introduced in the wave equations for the field variables \(\Omega^{(\pm)}(\textbf{r},\tau)\):
\begin{eqnarray} \label{eq: field equations}
\label{eq: field equations 1}
\left[\frac{\partial}{\partial z}\mp\frac{i}{2 k_0}\left(\frac{\partial^2}{\partial x^2}+\frac{\partial^2}{\partial y^2}\right)+\frac{\mu_s(\textbf{r},\tau)}{2}\right]\Omega_{s}^{(\pm)}(\textbf{r},\tau)
 \nonumber\\= 
\pm i\frac{3}{8\pi}\lambda^2_0\Gamma_{\text{rad.}} P^{(\pm)}_s(\textbf{r},\tau),
\end{eqnarray}
where \(\lambda_0\) is the wavelength associated with the carrier frequency \(\omega_0\), and the functions \(\mu_s(\textbf{r},\tau)\) represent the absorption coefficients that may depend on time since atomic populations change. The polarization fields \(P_s^{(\pm)}(\textbf{r},\tau)\) are defined as follows:
 \begin{subequations}
  \label{eq: field equations 2-3}
    \begin{align}
    \label{eq: field equations 2}
        P^{(+)}_s(\textbf{r},\tau)=n(\textbf{r}) \sum_{u,\, l}T_{lu,s} \rho_{ul}(\textbf{r},\tau)+f_s(\textbf{r},\tau),\\
        \label{eq: field equations 3}
        P^{(-)}_s(\textbf{r},\tau)=n(\textbf{r}) \sum_{u,\, l}T_{ul,s} \rho_{lu}(\textbf{r},\tau)+g_s(\textbf{r},\tau),
    \end{align}
    \end{subequations}
where, in addition to the expected contribution of the atomic coherences \(\rho_{ul}(\textbf{r},\tau)\) and \(\rho_{lu}(\textbf{r},\tau)\), we include noise terms \(f_s(\textbf{r},\tau)\) and \(g_s(\textbf{r},\tau)\), with the following correlation properties:
\begin{subequations}
\label{eq: f_s noise correlations}
\begin{align}
\langle f_s(\textbf{r},\tau)f_{s'}(\textbf{r}',\tau')\rangle&=\langle f_{s}^\dag(\textbf{r},\tau)f_{s'}^\dag(\textbf{r}',\tau')\rangle=0,\\
\langle f_{s}(\textbf{r},\tau)f_{s'}^\dag(\textbf{r}',\tau')\rangle&= \delta_{ss'}\delta(\textbf{r}-\textbf{r}')\delta(\tau-\tau').
\end{align}
\end{subequations}
The same stochastic characteristics apply to \(g_s(\textbf{r},\tau)\) and \(g_s^\dag(\textbf{r},\tau)\). The pair \(f_s(\textbf{r},\tau)\) and \(f_s^\dag(\textbf{r},\tau)\) is statistically independent from \(g_s(\textbf{r},\tau)\) and \(g_s^\dag(\textbf{r},\tau)\). For a more detailed discussion of the noise terms, their derivations, and in particular, the proof that they reproduce spontaneous emission, refer to Ref.~\cite{benediktovitch2023stochastic}. { When integrated along the $z$-axis, the noise terms should be understood in the Itô sense. In terms of the other dimensions — namely, the transverse direction and retarded time — the noise terms should be regarded as smooth functions with a correlation distance that is sufficiently small.}

If the noise terms are completely omitted, the equations remain unchanged under the exchange of the variables \(\Omega^{(+)}_s(\textbf{r},\tau) \rightleftarrows \Omega^{(-)*}_s(\textbf{r},\tau)\) and \(\rho_{pq}(\textbf{r},\tau) \rightleftarrows \rho^*_{qp}(\textbf{r},\tau)\). Unfortunately, this symmetry breaks when the noise terms are considered because \(f_s^\dag(\textbf{r},\tau) \neq g_s^{\dag *}(\textbf{r},\tau)\) and \(f_s(\textbf{r},\tau) \neq g_s^{*}(\textbf{r},\tau)\). The main consequence of this asymmetry is that certain stochastic trajectories may become divergent. For more details, refer to Sec. III C in \cite{Chuchurka2023compact}.

{Note that Eq.~(\ref{eq:atomic_equations_4}) includes noise terms linearly and quadratically dependent on the atomic variables $\rho_{pq}(\textbf{r},\tau)$. In the following section, we will demonstrate that the linear terms play a key role in the formation of spontaneous emission, ensuring that its intensity is proportional to the populations of the upper states. Generally, atoms produce both spontaneous and stimulated emission. Ignoring the quadratic terms, the total intensity turns out to be overestimated. This discrepancy, however, is anticipated to be negligible at high atomic concentrations. In Ref. \cite{benediktovitch2023stochastic}, the quadratic terms were discarded, in this work, they are retained. A detailed discussion can be found in Appendix \ref{Role of the quadratic noise terms}.

Careful observation reveals that the linear noise terms in Eq.~(\ref{eq:atomic_equations_4}) resemble the two terms in Eq.~(\ref{eq:atomic_equations_3}) in the square brackets that couple the atomic variables to the Rabi frequencies $\Omega_s^{(\pm)}(\textbf{r},\tau)$. These two terms describe the stimulated emission of light, while the remaining terms account for stimulated absorption. Since only emission processes can be spontaneous, there are only two linear noise terms in Eq.~(\ref{eq:atomic_equations_4}) corresponding to the terms responsible for stimulated emission.
}

We must highlight that the right-hand side of Eq.~(\ref{eq: field equations 1}) may include modes beyond the paraxial approximation due to white noise in the source. To resolve this, damping for non-paraxial modes must be added to Eq.~(\ref{eq: field equations 1}).

To conclude this section, we present the guidelines for constructing observables based on a set of realizations of stochastic variables. Here, we provide the key expressions, while additional details can be found in Ref.~\cite{benediktovitch2023stochastic}, particularly in Sec. III E.

To directly compute the population $p_q(\textbf{r},\tau)$ of state $\ket{q}$, we employ the following formula:
\begin{equation*}
p_q(\textbf{r},\tau)=\langle \rho_{qq}(\textbf{r},\tau)\rangle,
\end{equation*}
For analyzing the properties of emitted fields, we utilize the following expression to calculate intensities:
\begin{subequations}
\label{eq: observables for the fields}
\begin{equation}
\label{eq: observables for the fields a}
I_s(\textbf{r},\tau) = \hbar \omega_0 \frac{\langle \Omega^{(+)}_s(\textbf{r},\tau)\Omega^{(-)}_s(\textbf{r},\tau)\rangle}{\frac{3}{8\pi}\lambda_0^2\Gamma_{\text{rad.}}}.
\end{equation}
The spectrum is determined by:
\begin{equation}
\label{eq: observables for the fields b}
\!\!S_s(\textbf{r}, \omega) = \int \frac{d\tau d\tau'}{(2\pi)^2}\langle \Omega^{(+)}_s(\textbf{r},\tau)\Omega^{(-)}_s(\textbf{r},\tau')\rangle e^{i\omega(\tau-\tau')}.\!\!
\end{equation}
\end{subequations}
{Here, $s$ specifies the polarization.}

%%%%%%%%%%%%%%%%%%%%%%%%%%%%%%%%%%%%%%%%%%%%%%%%%%%%%%%%%
\section{Perturbative Treatment of Spontaneous Emission }\label{sec: linear}
%%%%%%%%%%%%%%%%%%%%%%%%%%%%%%%%%%%%%%%%%%%%%%%%%%%%%%%%%

{As mentioned in the Introduction, the stochastic differential equations outlined in Sec.~\ref{sec: stochastic formalism} exhibit instabilities. These instabilities arise from the fact that the noise components cause the stochastic variables to become non-Hermitian, as briefly explained in Sec.~\ref{sec: stochastic formalism}. Specifically, $\rho_{pq}(\textbf{r},\tau)\neq\rho_{qp}^*(\textbf{r},\tau)$, and similarly for the field variables, $\Omega_{s}^{(+)}(\textbf{r},\tau)\neq\Omega_{s}^{(-)*}(\textbf{r},\tau)$. To tackle the instability issue, we begin by simplifying the structure of the stochastic components of the Maxwell-Bloch equations. 

Rather than examining the atomic dynamics in the entire medium at once, we divide it into smaller regions and concentrate on the atomic dynamics within one of these regions. Each region is sufficiently small to assume that the incoming field is significantly stronger compared to the field produced within it. Consequently, we may assume that the Rabi frequencies $\Omega_s^{(\pm)}(\textbf{r},\tau)$ in Eq. (\ref{eq:atomic_equations_3}) only represent the light produced externally to the considered small region. These incoming fields, together with the noise terms $g_s^\dag(\textbf{r},\tau)$ and $f_s^\dag(\textbf{r},\tau)$, and incoherent processes in Eqs. (\ref{eq:atomic_equations_1}) and (\ref{eq:atomic_equations_2}), condition the atomic dynamics in the considered region. It determines polarization fields $P_{s}^{(\pm)}(\textbf{r},\tau)$ in the region, which serve as a source in the field equations (\ref{eq: field equations}) modifying the fields passing through the region.

Depending on the intensity of the incoming fields, the atoms can exhibit two distinct qualitative behaviors. If the fields are weak and largely consist of spontaneous emission, stochastic contributions play a pivotal role. In the opposite scenario in which the analyzed region is exposed to strong incoming fields, they primarily define the atomic dynamics, and the noise terms can be safely neglected. In other words, the problem simplifies to solving deterministic Maxwell-Bloch equations. 

Unlike the deterministic scenario, the atomic dynamics in the limit of weak fields is generally complex and can even exhibit diverging behavior, as mentioned at the beginning of this section. However, we note that the noise terms are only responsible for generating spontaneous photons that propagate within the sample. Upon leaving the considered region, they further influence the evolution of atoms in neighboring regions. {The rest of the spontaneous photons exiting the sample sideways constitutes the majority of the produced spontaneous emission, and merely gives rise to spontaneous-emission decay rates of the excited atoms.} Consequently, compared to the effect of the decay rates, the impact of the noise terms on the dynamics of the atomic variables can be treated perturbatively.

To maintain the Hermiticity of the dynamic variables in equations, it is crucial that the Rabi frequency of the electromagnetic fields remains significantly smaller than the total decay rate of the excited states. Given that strong incoherent processes are often present in x-ray superfluorescence experiments, this approximation is well-justified and establishes the quantitative limit for weak fields.

\subsection*{Strong Field Limit}

We begin with a less complicated case assuming that a given small region is exposed to strong incoming fields. In this scenario, the atomic dynamics can be fully characterized by the deterministic Bloch equations, namely, Eqs. (\ref{eq:atomic_equations_1}) -- (\ref{eq:atomic_equations_3}) in the presence of incoming fields $\Omega_s^{(\pm)}(\textbf{r},\tau)$ and incoherent processes but without any stochastic contributions. 

The solution to these equations is a deterministic density matrix denoted $\rho_{pq, \text{det.}}(\textbf{r},\tau)$, which can be used to infer various atomic properties. Specifically, the coherences between upper and lower states are particularly important. They are explicitly included in the polarization fields $P_s^{(\pm)}(\textbf{r},\tau)$, and they serve as a source of emission that propagates further through the medium, influencing atomic dynamics outside the considered region. Neglecting any noise contributions, the polarization fields $P_{s,\text{det.}}^{(\pm)}(\textbf{r},\tau)$ based solely on the deterministic density matrix $\rho_{pq, \text{det.}}(\textbf{r},\tau)$ have the following form:
 \begin{align*}
        P^{(+)}_{s,{\text{det.}}}(\textbf{r},\tau)=n(\textbf{r}) \sum_{u,\, l}T_{lu,s} \rho_{ul,{\text{det.}}}(\textbf{r},\tau),\\
        \label{eq: field equations 3}
        P^{(-)}_{s,{\text{det.}}}(\textbf{r},\tau)=n(\textbf{r}) \sum_{u,\, l}T_{ul,s} \rho_{lu,{\text{det.}}}(\textbf{r},\tau).
    \end{align*}

Apart from the limit of strong fields, the deterministic solution can also be used further in the perturbative analysis of the dynamics driven by weak fields.

\subsection*{Weak Field Limit}

In situations where the incoming fields are weak and primarily consisting of spontaneous emission, the noise terms  $f_s^\dag(\textbf{r},\tau)$, $g_s^\dag(\textbf{r},\tau)$ in the Bloch equations (\ref{eq:atomic_equations}), and $f_s(\textbf{r},\tau)$, $g_s(\textbf{r},\tau)$ in the field equations (\ref{eq: field equations}) cannot be neglected. However, assuming that most of the spontaneous emission escapes the sample by propagating sideways, the noise terms representing the remaining spontaneous photons can be treated perturbatively when compared to the effect of the decay rates. As a result, the previously estimated deterministic polarization fields $P^{(\pm)}_{s, \text{det.}}(\textbf{r},\tau)$  will be enhanced with additional stochastic corrections $P^{(\pm)}_{s, \text{noise}}(\textbf{r},\tau)$:
\begin{equation*}
    P^{(\pm)}_{s}(\textbf{r},\tau)=P^{(\pm)}_{s, \text{det.}}(\textbf{r},\tau)+P^{(\pm)}_{s, \text{noise}}(\textbf{r},\tau).
\end{equation*} }

{We integrate the Bloch equations (\ref{eq:atomic_equations}) for the coherences $\rho_{ul}(\textbf{r},\tau)$ and $\rho_{lu}(\textbf{r},\tau)$, retaining the noise terms $g^\dag_s(\textbf{r},\tau)$ and $f^\dag_s(\textbf{r},\tau)$ only up to a first-order perturbation. The resulting expressions can be simplified by assuming that the rates $\Gamma_u+\Gamma_l$ significantly exceed the Rabi frequency $\Omega^{(\pm)}(\textbf{r},\tau)$ {and energy splittings $\Delta\omega_{uu'}$ and $\Delta\omega_{ll'}$}.  This approximation is crucial as it enables us to provide simple expressions that preserve Hermiticity of the dynamic variables. Inserting the result in the expressions for the polarization fields $P^{(\pm)}_{s}(\textbf{r},\tau)$ in Eq.~(\ref{eq: field equations 2-3}), we obtain the stochastic corrections $P^{(\pm)}_{s, \text{noise}}(\textbf{r},\tau)$ to the previously estimated deterministic polarization fields $P^{(\pm)}_{s, \text{det.}}(\textbf{r},\tau)$.
The explicit expression for the stochastic contribution $P^{(+)}_{s, \text{noise}}(\textbf{r},\tau)$ reads:}
\begin{multline}
\label{eq: raw stochastic polarization 1}
    P^{(+)}_{s,\text{noise}}(\textbf{r},\tau)\approx f_s(\textbf{r},\tau)+n(\textbf{r})\int_{0}^{\tau}d\tau'e^{(\Gamma_u+\Gamma_l)(\tau'-\tau)/2}\\\times\sum_{s'}\rho_{ss'}^{\text{(up.)}}(\textbf{r},\tau') g_{s'}^\dag(\textbf{r},\tau'),
\end{multline}
where {the noise term $f_s(\textbf{r},\tau)$ originates directly from Eq.~(\ref{eq: field equations 2-3}), and the remaining terms are derived from the expressions for the coherences $\rho_{ul}(\textbf{r},\tau)$. To shorten the expression, }we have introduced an effective density matrix for the upper states $\rho_{ss'}^{\text{(up.)}}(\textbf{r},\tau)$, defined as follows:
\begin{multline}
\label{eq: effective upper state population}
    \rho_{ss'}^{\text{(up.)}}(\textbf{r},\tau)=\sum_{\mathclap{u',u,l}}T_{lu,s}{\rho}_{uu'\!, {\text{det.}}}(\textbf{r},\tau)T_{u'l,s'}\\-\sum_{l,u}\rho_{ul, {\text{det.}}}(\textbf{r},\tau)T_{lu,s}\sum_{l',u'}\rho_{l'u'\!, {\text{det.}}}(\textbf{r},\tau)T_{u'l',s'}.
\end{multline}
Likewise, for $P_{s,\text{noise}}^{(-)}(\textbf{r},\tau)$ we find:
\begin{multline}
\label{eq: raw stochastic polarization 2}
    P^{(-)}_{s,\text{noise}}(\textbf{r},\tau)\approx g_s(\textbf{r},\tau)+n(\textbf{r})\int_{0}^{\tau}d\tau'e^{(\Gamma_u+\Gamma_l)(\tau'-\tau)/2}\\\times\sum_{s'} f_{s'}^\dag(\textbf{r},\tau')\rho_{s's}^{\text{(up.)}}(\textbf{r},\tau').
\end{multline}

{The expressions of the stochastic polarization fields $P_{s,\text{noise}}^{(\pm)}(\textbf{r},\tau)$ in Eqs. (\ref{eq: raw stochastic polarization 1}) and (\ref{eq: raw stochastic polarization 2}) involve integration over the retarded time starting from zero. This signifies the initial moment in time for which the initial conditions are given.}

Due to the correlation properties of the noise, the stochastic polarizations fields $P^{(\pm)}_{s,\text{noise}}(\textbf{r},\tau)$ exhibit zero means. As they are linearly dependent on the noise terms $f(\textbf{r},\tau)$, $g(\textbf{r},\tau)$, $f^\dag(\textbf{r},\tau)$ and $g^\dag(\textbf{r},\tau)$, their statistical characteristics obey a Gaussian distribution. It is sufficient to list their second-order correlation functions to fully characterize their statistics:
\begin{subequations}
\label{eq: spontaneous correlator}
\begin{multline}
    \langle  P_{s,\text{noise}}^{(+)}(\textbf{r},\tau) P_{s',\text{noise}}^{(+)}(\textbf{r}',\tau')\rangle\\=\langle  P_{s,\text{noise}}^{(-)}(\textbf{r},\tau) P_{s',\text{noise}}^{(-)}(\textbf{r}',\tau')\rangle=0,
\end{multline}
\begin{multline}
\label{eq: spontaneous correlator b}
    \langle  P_{s,\text{noise}}^{(+)}(\textbf{r},\tau) P_{s',\text{noise}}^{(-)}(\textbf{r}',\tau')\rangle \\= n(\textbf{r})e^{-(\Gamma_u+\Gamma_l)|\tau-\tau'|/2}\rho_{ss'}^{\text{(up.)}}(\textbf{r},\min[\tau,\tau'])\delta(\textbf{r}-\textbf{r}').
\end{multline}
\end{subequations}
{The correlator in Eq.~(\ref{eq: spontaneous correlator b}) is proportional to the effective density matrix for the upper states $\rho_{ss'}^{\text{(up.)}}(\textbf{r},\tau)$ defined in Eq.~(\ref{eq: effective upper state population}). This effective density matrix includes terms that depend linearly and quadratically on the atomic variables $\rho_{pq}(\textbf{r},\tau)$, originating directly from the linear and quadratic stochastic contributions in Eq.~(\ref{eq:atomic_equations_4}). The linear terms capture the essential property of spontaneous emission, making its intensity proportional to the population of the upper states.\footnote{Indeed, in Ref. \cite{benediktovitch2023stochastic}, where quadratic noise terms are disregarded, these linear terms are present in the first-order correlation function of spontaneous emission (see Sec. III F in \cite{benediktovitch2023stochastic}).} On the other hand, the quadratic terms provide a quantitative correction, which becomes negligible at high atomic concentrations. A detailed illustration can be found in Appendix \ref{Role of the quadratic noise terms}. }

Apart from $P_{s,\text{noise}}^{(\pm)}(\textbf{r},\tau)$, there are no other stochastic objects affecting the dynamics of the fields $\Omega_s^{(\pm)}(\textbf{r},\tau)$. The right-hand side of the wave equation~(\ref{eq: field equations}) contains only the stochastic polarization fields $P_{s,\text{noise}}^{(\pm)}(\textbf{r},\tau)$ and the atomic variables $\rho_{pq,\text{det.}}(\textbf{r},\tau)$. Although the number of independent noise terms is reduced, direct sampling of $P^{(\pm)}_{s,\text{noise}}(\textbf{r},\tau)$ through the diagonalization of the correlator for each $\textbf{r}$-coordinate is computationally intractable. Fortunately, it is possible to derive a compact representation that not only enhances computational efficiency, but, more importantly, ensures that $P_{s,\text{noise}}^{(+)}(\textbf{r},\tau)$ is the complex conjugate of $P_{s,\text{noise}}^{(-)}(\textbf{r},\tau)$, thereby guaranteeing numerical stability. The proposed representation is introduced in the next section.

Additionally, it is important to note that $\rho_{pq,\text{det.}}(\textbf{r},\tau)$ can fully replace the role of $\rho_{pq}(\textbf{r},\tau)$ when constructing the expectation values (refer to the end of Sec. \ref{sec: stochastic formalism}) for the atomic properties. Indeed, the original stochastic Bloch equations only contain uncorrelated noise terms $f_s^\dag(\textbf{r},\tau)$ and $g_s^\dag(\textbf{r},\tau)$. Consequently, their absence is unnoticeable when focusing exclusively on atomic expectation values. Based on this justification, we remove the subscript $_{\text{det.}}$ from the atomic variables in the following sections.

{Finally, we observe that the equations for both limits of weak and strong fields share a similar structure. Atomic dynamics is described by deterministic Bloch equations. The key distinction lies in the treatment of weak fields, where inclusion of stochastic polarization fields in the field equations becomes necessary to address spontaneous emission. This correction is only suitable for the weak-field regime. However, when the Rabi frequency of the electromagnetic fields become comparable to the inverse total lifetimes, even though the stochastic correction is redundant, its contribution is negligible compared to the strong incoming fields. Hence, we can incorporate this correction in both scenarios discussed in this section.

For the sake of simplicity, we analyzed the atomic behavior within small regions. However, the resulting equations, being applicable to each individual region, hold true for the entire medium.}

%%%%%%%%%%%%%%%%%%%%%%%%%%%%%%%%%%%%%%%%%%%%%%%%%%%%%%%%%
\section{Self-consistent system of equations}\label{sec: self}
%%%%%%%%%%%%%%%%%%%%%%%%%%%%%%%%%%%%%%%%%%%%%%%%%%%%%%%%%

{In Sec. \ref{sec: linear},  we took a step toward addressing the stability issue discussed in the Introduction and simplified the structure of the stochastic components. While we outlined their general statistical properties, they are unsuitable for efficient numerical computations. Here, we present a system of equations that can be solved efficiently and, crucially, maintain the Hermiticity of the dynamic variables.} 

%We made the assumption that noise terms could be disregarded when their impact is dominated by the stimulation processes induced by strong incoming fields. Conversely, in cases where the fields are weak and largely consisting of spontaneous emission, the noise terms are preserved but treated perturbatively. This is based on the assumption that incoherent processes are sufficiently strong and primarily define the atomic dynamics. }

Based on the derivations in Sec. \ref{sec: linear}, we conclude that the Bloch equations (\ref{eq:atomic_equations_1}) -- (\ref{eq:atomic_equations_3}), excluding the noise terms in Eq.~(\ref{eq:atomic_equations_4}), effectively describe the atomic degrees of freedom. {The noise terms in Eq.~(\ref{eq:atomic_equations_4}) have been relocated into the wave equations.} The general form of the wave equations remains unchanged as in Eq.~(\ref{eq: field equations 1}):
\begin{eqnarray*}
\left[\frac{\partial}{\partial z}\mp\frac{i}{2 k_0}\left(\frac{\partial^2}{\partial x^2}+\frac{\partial^2}{\partial y^2}\right)+\frac{\mu_s(\textbf{r},\tau)}{2}\right]\Omega_{s}^{(\pm)}(\textbf{r},\tau)
 \nonumber\\= 
\pm i\frac{3}{8\pi}\lambda^2_0\Gamma_{\text{rad.}} P^{(\pm)}_s(\textbf{r},\tau).
\end{eqnarray*}
However, the expressions for the polarization fields in Eqs.~(\ref{eq: field equations 2}) and (\ref{eq: field equations 3}) are modified:
\begin{subequations}
\label{eq: full P with noise}
\begin{align}
        \!\!P^{(+)}_s(\textbf{r},\tau)=n(\textbf{r}) \sum_{u,\, l}T_{lu,s} \rho_{ul}(\textbf{r},\tau)+ P^{(+)}_{s,\text{noise}}(\textbf{r},\tau),\!\!\\
        \!\!P^{(-)}_s(\textbf{r},\tau)=n(\textbf{r}) \sum_{u,\, l}T_{ul,s} \rho_{lu}(\textbf{r},\tau)+ P^{(-)}_{s,\text{noise}}(\textbf{r},\tau).\!\!
\end{align}
\end{subequations}
Here, the deterministic coherences $\rho_{ul}(\textbf{r},\tau)$ are accompanied by the noise polarization fields $P^{(\pm)}_{s,\text{noise}}(\textbf{r},\tau)$. {At this point, these noise polarization fields not only include noise terms $f_s(\textbf{r},\tau)$ and $g_s(\textbf{r},\tau)$, but they also incorporate the effect of the atomic noise terms in Eq.~(\ref{eq:atomic_equations_4}), as demonstrated in Sec.~\ref{sec: linear}}. It is important to note that the correlation properties in Eq.~(\ref{eq: spontaneous correlator}) conditioning $P^{(\pm)}_{s,\text{noise}}(\textbf{r},\tau)$ do not uniquely define a specific form of the stochastic polarization fields. Nevertheless, assuming they should be complex conjugates of each other resolves this ambiguity. {This marks the final step in formulating the stochastic equations free of diverging solutions.} The stochastic polarization fields are found as a solution to the following equation:
\begin{multline}
\label{eq: stochastic polarization equation}
    \frac{\partial}{\partial \tau} P_{s,\text{noise}}^{(+)}(\textbf{r},\tau)=-(\Gamma_u+\Gamma_l) P_{s,\text{noise}}^{(+)}(\textbf{r},\tau)/2\\+\sum_{s'} \xi_{ss'}(\textbf{r},\tau)F_{s'}(\textbf{r},\tau).
\end{multline}
Notably, the stochastic polarizations share the same decay rate $(\Gamma_u+\Gamma_l)/2$ with $\rho_{ul}(\textbf{r},\tau)$. The source term consists of Gaussian white noise terms $F_{s}(\textbf{r},\tau)$ with zero means and the following correlation properties:
\begin{subequations}
\label{eq: F correlator}
\begin{align}
    \langle F_{s}^*(\textbf{r},\tau)F_{s'}(\textbf{r}',\tau')\rangle &= \delta_{ss'}\delta(\textbf{r}-\textbf{r}')\delta(\tau-\tau'),\\
    \langle F_{s}(\textbf{r},\tau)F_{s'}(\textbf{r}',\tau')\rangle &= 0.
\end{align}
\end{subequations}
To restore the correct statistics given by Eq.~(\ref{eq: spontaneous correlator}), these elementary noise terms are multiplied by the functions $\xi_{ss'}(\textbf{r},\tau)$, obtained from the following diagonalization problem:
\begin{multline}
\label{eq: diagonalization 1}
    \sum_i \xi_{si}(\textbf{r},\tau)\xi_{s'i}^*(\textbf{r},\tau)\\= n(\textbf{r})\left[\frac{\partial}{\partial \tau}+\Gamma_u+\Gamma_l\right]{\rho}^{(\text{up.})}_{ss'}(\textbf{r},\tau),
\end{multline}
which features the effective upper state population ${\rho}^{(\text{up.})}_{ss'}(\textbf{r},\tau)$ defined as follows:

{\begin{multline}
\label{eq: effective upper state population 1}
    \rho_{ss'}^{\text{(up.)}}(\textbf{r},\tau)=\sum_{\mathclap{u',u,l}}T_{lu,s}{\rho}_{uu'}(\textbf{r},\tau)T_{u'l,s'}\\-\sum_{l,u}\rho_{ul}(\textbf{r},\tau)T_{lu,s}\sum_{l',u'}\rho_{l'u'}(\textbf{r},\tau)T_{u'l',s'}.
\end{multline}}
Comparing with the phenomenological approach in Ref. \cite{2000'Larroche}, we notice two significant differences. First, the noise terms are not directly included in the equations for the atomic variables. Second, the multipliers $\xi_{ss'}(\textbf{r},\tau)$ depend not only on the populations of the upper states but also on their derivatives. 
In instances in which the upper states are instantly populated, we observe the third important difference: non-zero initial conditions for $P^{(\pm)}_s(\textbf{r},0)$ must be included. Similarly to the source in Eq.~(\ref{eq: stochastic polarization equation}), the initial condition must possess certain statistics:
\begin{equation*}
P^{(+)}_s(\textbf{r},0)=\sum_{s'}\zeta_{ss'}(\textbf{r}) H_{s'}(\textbf{r}),
\end{equation*}
where $H_{s}(\textbf{r})$ follow Gaussian distributions with zero means and the following second-order correlators:
\begin{align*}
    \langle H_{s}^*(\textbf{r})H_{s'}(\textbf{r}')\rangle &= \delta_{ss'}\delta(\textbf{r}-\textbf{r}'),\\
    \langle H_{s}(\textbf{r})H_{s'}(\textbf{r}')\rangle &= 0.
\end{align*}
Similarly to $\xi_{ss'}(\textbf{r},\tau)$, the functions $\zeta_{ss'}(\textbf{r})$ are obtained from the following diagonalization problem:
\begin{equation}
\label{eq: diagonalization 2}
     \sum_i \zeta_{si}(\textbf{r})\zeta_{s'i}^*(\textbf{r}) = n(\textbf{r}){\rho}^{(\text{up.})}_{ss'}(\textbf{r},0).
\end{equation}
The obtained random initial conditions are reminiscent of the methodology from Refs. \cite{Gross1982, 1979'Vrehen}, where similar random initial conditions have been used to describe superfluorescence in the case of instant pumping and $\Gamma_l = 0$.

Finally, let us point out that the proposed representation for the stochastic polarization fields exists only when the diagonalization problems in Eqs. (\ref{eq: diagonalization 1}) and (\ref{eq: diagonalization 2}) have solutions. There are no solutions when, for example, the effective upper state population changes so fast that the right-hand side of Eq.~(\ref{eq: diagonalization 1}) becomes negative. This limitation, however, arises only when the fields become so intense that the Rabi amplitudes become comparable to the inverse lifetimes of the states. However, under such circumstances, the contribution of spontaneous emission becomes negligible and the noise terms can be omitted. In Sec.~\ref{sec: stochastic formalism}, we imposed the restriction that states associated with the same manifold must possess identical lifetimes. This constraint guarantees solutions of the diagonalization problems Eqs. (\ref{eq: diagonalization 1}) and (\ref{eq: diagonalization 2}).

%%%%%%%%%%%%%%%%%%%%%%%%%%%%%%%%%%
\section{One-dimensional treatment and numerical examples}\label{numerical examples}
%%%%%%%%%%%%%%%%%%%%%%%%%%%%%%%%%%
 \begin{figure}[t!]
    \centering
    \includegraphics[width = 0.65\linewidth]{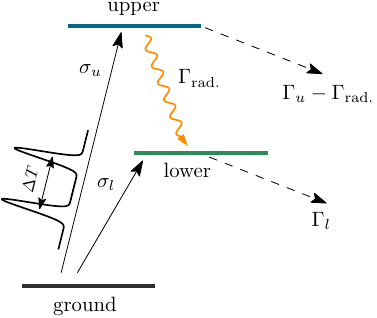}
    \caption{Level scheme used in the numerical simulations with the following parameters, adapted to the case of the Cu-like K-$\alpha$ transition of $\omega_0=8.048$ {keV}: decay rates $\Gamma_u=2.1$ fs$^{-1}$ and $\Gamma_l=0.9$ fs$^{-1}$, the rate of spontaneous emission $\Gamma_{\text{rad.}}=0.6$ fs$^{-1}$, photoionization cross-sections $\sigma_u=25.3$ Kb and $\sigma_l=9.5$ Kb. A fraction of this fluorescence (this fraction is defined in Sec.~\ref{Superfluorescence}), travels through the sample, inducing superfluorescence. In reality, $\sigma_l$ is considerably lower for Cu. We have increased this parameter to induce more dynamics to the lower state. The atomic concentration is assumed as $n=85.0$ nm$^{-3}$.}
    \label{fig: energy levels}
\end{figure}

To illustrate the functioning of the noise terms, we present the simplest example demonstrating superfluorescent behavior, considering a system of three-level atoms that exhibit properties akin to Cu K-$\alpha$ transition (see Fig. \ref{fig: energy levels} for parameters). Although the provided formalism is suitable for paraxial geometry, we neglect diffraction and assume that light propagates strictly parallel along the medium. Specifically, we analyze the dynamics at the transverse center of the pump pulse distribution, with the assumption of single-polarization fields.

Typically, x-ray superfluorescence is observed from x-ray transitions opened through rapid photoionization by x-ray Free Electron Lasers (XFELs) \cite{Rohringer2012, Yoneda2015, Duguay1967, Kapteyn1992, Kimberg2013}. XFEL pulses, operating in Self-Amplified Spontaneous Emission (SASE) mode, exhibit characteristic spiky spectral and {temporal} profiles. Such a pulse comprises multiple attosecond sub-pulses with random intensities \cite{Saldin2008, Saldin1998} whose numerical models may be found in Refs. \cite{Cavaletto2021, Pfeifer2010}. Recent advancements enable the production of single attosecond bursts of x-ray radiation \cite{Marinelli2024}. Moreover, schemes and split-and-delay lines are developed to provide pairs of attosecond pulses.
%%%%%%%%%%%%%%%%%%%%%%%%%%%%%%%%%%%%%%%%%%%%%%%%%%%%%%%%%%%%%%%%
 \begin{figure*}[t!]

    \includegraphics[width = \linewidth]{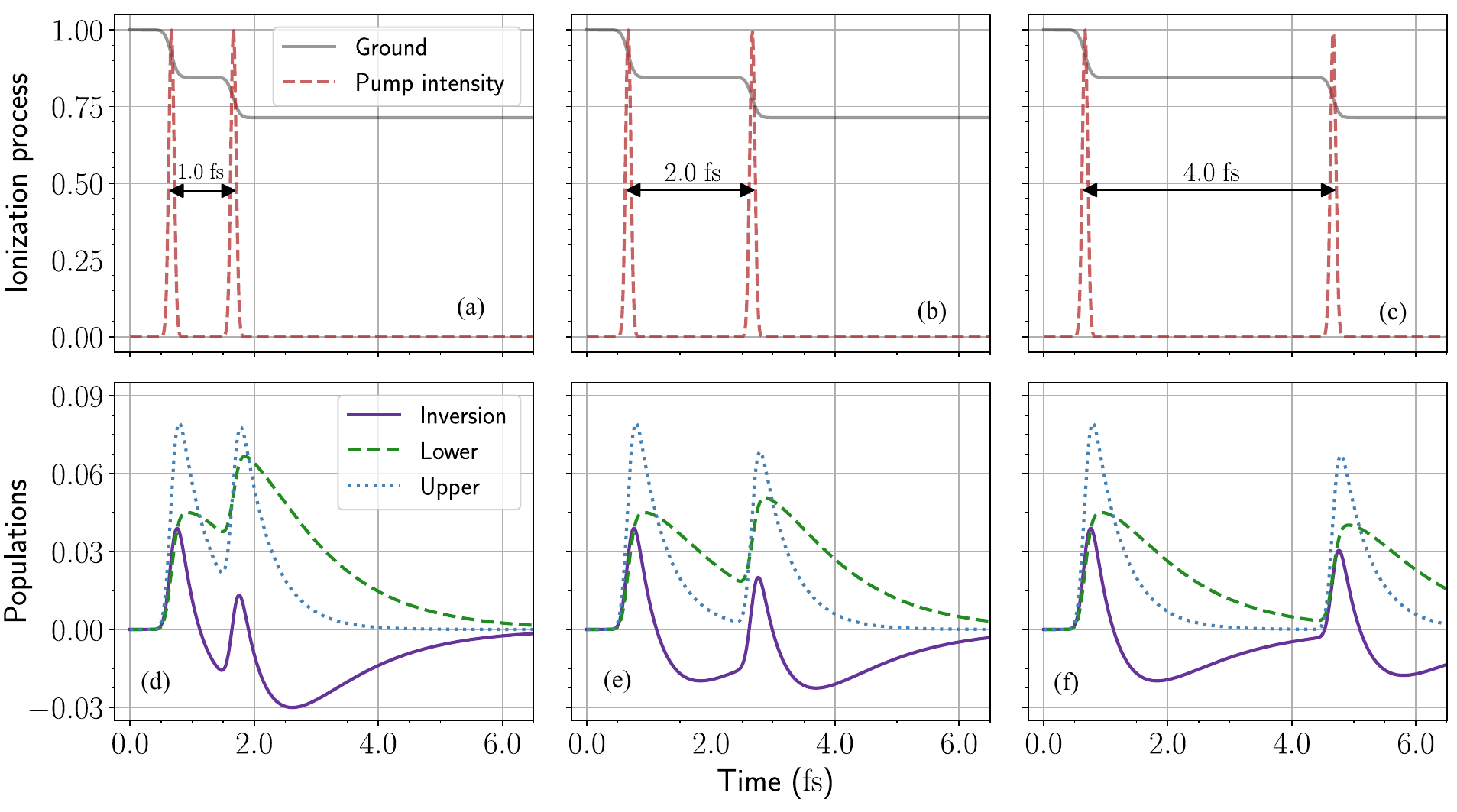}
    \caption{Dynamics induced by a pump consisting of two successive sub-pulses. Spontaneous emission and the ensuing superfluorescence are disregarded. Each column represents a distinct time delay between the sub-pulses, specifically $\Delta T = 1.0$ fs, $2.0$ fs, $4.0$ fs. As we neglect pump absorption, our analysis focuses solely on the dynamics of a single atom. Panels (a) to (c) showcase normalized pump intensity profiles. The pump pulses induce photoionization, depleting the ground state population, as illustrated in (a) to (c). Photoionization promotes atoms to excited states, and due to finite lifetimes, these states decay further, as depicted in (d) to (f). Additionally, the solid purple curve denotes the population inversion between the upper and lower excited states.}
    \label{fig: populations}
\end{figure*}
%%%%%%%%%%%%%%%%%%%%%%%%%%%%%%%%%%%%%%%%%%%%%%%%%%%%%%%%%%%%%%%%

While accurately modeling specific pumping schemes can lead to complex atomic evolution, potentially obscuring the demonstration, our goal is to capture the key qualitative properties of the pumping process. Hence, we assume that the pumping emission consists of two sub-pulses with variable time delays (set to $1.0$, $2.0$, and $4.0$ fs for demonstration), each with a full-width at half-maximum (FWHM) intensity duration of $0.2$ fs and a peak intensity of $3.2\times10^{19}$ W/cm$^2$. These parameters correspond to $1.6$ \textmu J or $1.1\times 10^9$ photons per sub-pulse, with a focusing of $200\times 100$~nm$^2$ and $9.0$ {keV} photon energy, typically achievable at existing XFEL sources. The pulse intensity $I(\tau)$ is numerically represented by a sum of two Gaussian functions. In our simulations, we disregard the absorption of the XFEL pump pulse.

We compare our method with simulations based on a commonly used approach employing phenomenological noise terms \cite{2000'Larroche, 2018'Krusic}. In the numerical simulations, we discretize variables with a time step of 1.0 as and a coordinate step of $3.5\times10^{-2}$ \textmu m. The time integration of the deterministic components of the atomic variables is conducted using the Runge-Kutta fourth-order method, implemented in the \texttt{DifferentialEquations.jl} library \cite{Rackauckas2017}. For the deterministic parts of the field variables, we employ the trapezoid rule. The integration of the noise terms is carried out using the Euler-Maruyama method. The statistical sample comprises $2400$ realizations; for the diagram in Fig.~\ref{fig: phases} (g), we have used $8000$ realizations.

%%%%%%%%%%%%%%%%%%%%%%%%%%%%%%%%%%%%%%%%%%%
\subsection{Preparation of the population inversion}
%%%%%%%%%%%%%%%%%%%%%%%%%%%%%%%%%%%%%%%%%%%
The depletion of the ground state is governed by the rate equation:
\begin{subequations}
\label{eq: rate equations for the pumping}
\begin{equation}
    \frac{\partial}{\partial \tau}\rho^{(\text{ground})}(z,\tau)=-\sigma I(\tau) \rho^{(\text{ground})}(z,\tau),
\end{equation}
where $\sigma = \sigma_u+\sigma_l$ is the total photoionization cross-section, which populates the upper and lower ionic states as follows:
\begin{align}
\begin{multlined}
    \frac{\partial}{\partial \tau}\rho_{uu}(z,\tau) =-\Gamma_u\rho_{uu}(z,\tau)\quad\quad\quad\quad\quad\,\\+\sigma_u I(\tau) \rho^{(\text{ground})}(z,\tau),
\end{multlined}\\
\begin{multlined}
\frac{\partial}{\partial \tau}\rho_{ll}(z,\tau) =\Gamma_{\text{rad.}}\rho_{uu}(z,\tau)-\Gamma_l\rho_{ll}(z,\tau)\\
+\sigma_l I(\tau) \rho^{(\text{ground})}(z,\tau).
\end{multlined}
\end{align}
\end{subequations}
Superfluorescence is conditioned by the population inversion $\rho_{uu}(z,\tau)-\rho_{ll}(z,\tau)$, which, along with the upper- and lower-state populations, is depicted in Fig.~\ref{fig: populations} for the entrance of the gain medium. In the initial stages, when superfluorescence has not yet developed, the level populations are conditioned by the pump rate, lifetimes, and isotropic fluorescence. Consequently, the temporal dynamics of the populations remain approximately uniform across the sample. Only at the onset of saturation of the superfluorescent emission (non-linear regime) is the emitted field strong enough to alter the populations. Consequently, spontaneous emission traveling along the sample is neglected in Figs.~\ref{fig: populations} (d) to (f). In the subsequent simulations, it is properly taken into account.

%%%%%%%%%%%%%%%%%%%%%%%%%%%%%%%%%%%%%%
\subsection{Superfluorescence}
\label{Superfluorescence}
%%%%%%%%%%%%%%%%%%%%%%%%%%%%%%%%%%%%%%
Let us analyze the dynamics of spontaneous emission and its subsequent amplification. The evolution of the emitted field is determined by:
\begin{multline}
\label{eq: field 1d}
\frac{\partial}{\partial z}\Omega^{(+)}(z,\tau)
 \\=i\frac{3}{8\pi}\lambda^2_0\Gamma_{\text{rad.}} \left[
n\rho_{ul}(z,\tau)+ 
P_{\text{noise}}^{(+)}(z,\tau)
\right],
\end{multline}
with the spontaneous emission being determined by the stochastic polarization fields \(P_{\text{noise}}^{(+)}(z,\tau)=P_{\text{noise}}^{(-)*}(z,\tau)\):
\begin{equation}
\label{eq: one-dimensional field}
    \frac{\partial}{\partial \tau} P_{\text{noise}}^{(+)}(z,\tau)=- \Gamma P_{\text{noise}}^{(+)}(z,\tau)/2+\xi(z,\tau)F(z,\tau),
\end{equation}
where $\Gamma$ represents the sum of the inverse lifetimes of the upper and lower states and defines the timescales of the superfluorescence:
\begin{equation*}
    \Gamma=\Gamma_u+\Gamma_l.
\end{equation*}
The characteristic length scale of the amplification is governed by the gain length, given by:
\begin{equation*}
    L_g = \frac{8\pi}{3} \frac{\Gamma}{\Gamma_{\text{rad.}}}\left[\Delta Pn\lambda_0^2  \right]^{-1},
\end{equation*}
where \(\Delta P\) represents the effective population inversion (defined as half of the maximum value of the temporal population inversion). For our parameters,  $\Delta P \approx 0.02$ and $L_g \approx 1.16$ ~\textmu m.

%%%%%%%%%%%%%%%%%%%%%%%%%%%%%%%%%%%%%%%%%%%%%%%%%%%%%%%%
 \begin{figure*}[t!]
    \centering
    \includegraphics[width = \linewidth]{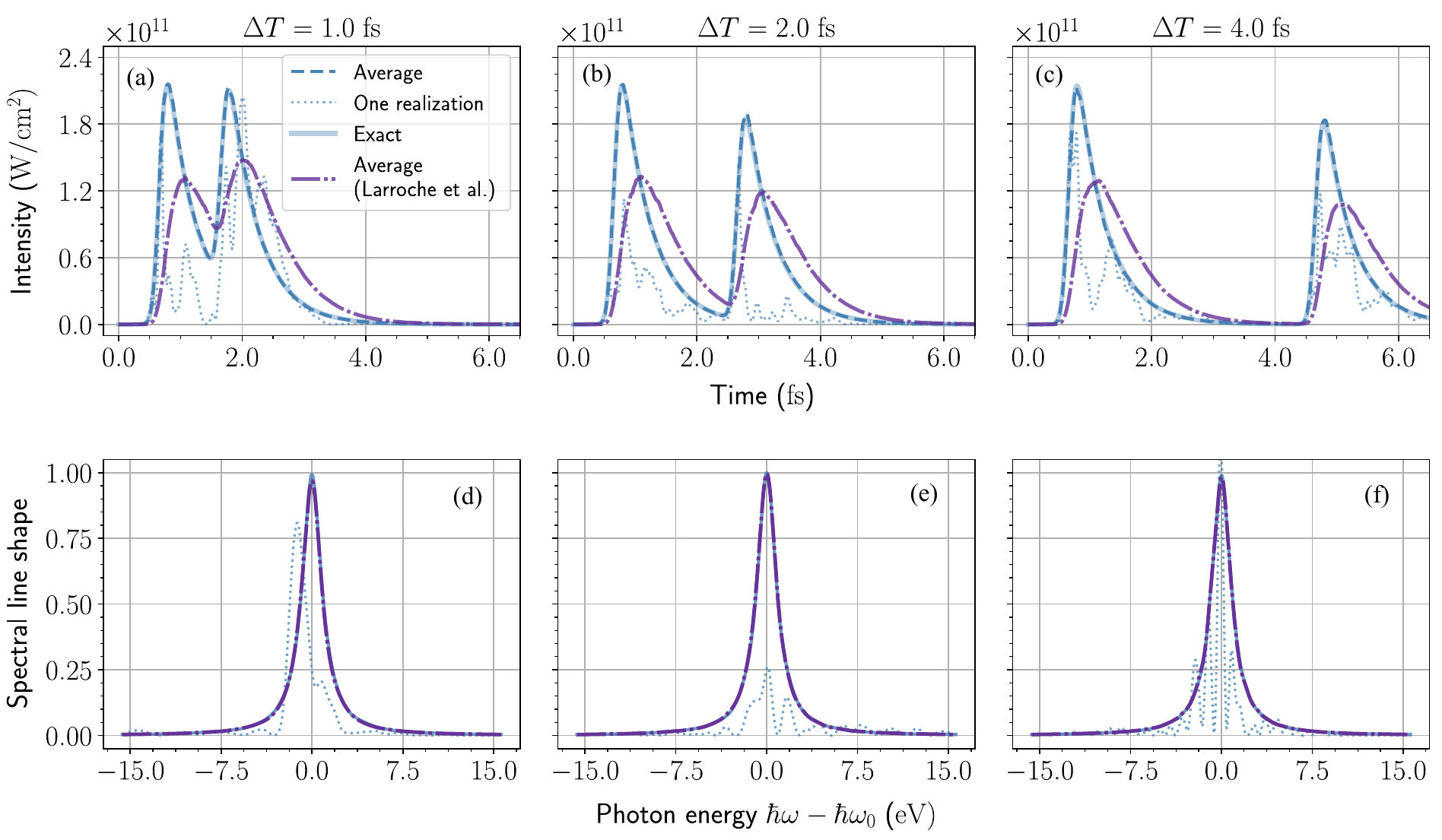}
    \caption{Temporal (a-c) and normalized spectral (d-f) intensities of spontaneous emission at the beginning of the sample $z=0.03L_g$ ($35.0$~nm) for separations of $\Delta T = 1.0$ fs, $2.0$ fs, $4.0$ fs between the two pump pulses. The solid blue line represents the exact average intensity based on Eq.~(\ref{eq: exact sp}), perfectly coinciding with the dashed blue line based on our methodology. The dotted blue line shows the intensity calculated based on an example of a single realization. The dash-dotted purple line provides the average intensity obtained from the phenomenological noise terms. In contrast to the time-dependent intensity profiles, the spectral intensities agree in all three methods.}
    \label{fig: spectra and intensities for first z}
\end{figure*}
%%%%%%%%%%%%%%%%%%%%%%%%%%%%%%%%%%%%%%%%%%%%%%%%%%%%%%%%%%%%%%%%

Care must be taken when introducing noise terms in a one-dimensional approximation. As suggested by Eq.~(\ref{eq: F correlator}), spontaneous emission is uncorrelated along the transverse direction. In terms of wave vectors, spontaneous emission is represented by wave vectors traveling in all possible directions. Those that violate the paraxial approximation must be neglected. However, the remaining wave vectors do not entirely contribute to the amplification process. If the field diffracts out of the region of positive population inversion before it is noticeably amplified, we can safely neglect it. The relevant wave vectors of spontaneous emission are localized within a small solid angle \(\Delta o\), corresponding to the first Fresnel zone defined by the gain length \(L_g\), namely \(\Delta o = \lambda_0/L_g\). This Fresnel zone covers an area \(S_{\text{coh.}} \sim \lambda_0^2/\Delta o \sim \lambda_0 L_g\). In this area, light is assumed to be transversely correlated. These assumptions constitute the one-dimensional approximation, which captures only the dynamics of the fields inside such areas of coherence. The noise terms consequently have the following correlation properties:
\begin{align*}
    \langle F(z,\tau) F^*(z',\tau')\rangle &= \frac{1}{S_\text{coh.}}\delta(\tau-\tau')\delta(z-z'),\\
    \langle F(z,\tau) \,F(z',\tau')\rangle&=0.
\end{align*}
In our numerical example, the value of the solid angle \(\Delta o\) is set to \(10^{-4}\), corresponding to a coherent area \(S_{\text{coh.}}\) of \(240\) nm\(^2\) --- roughly one-eightieth of the pump focus.

As the fields are assumed to have only one polarization, the diagonalization problem in Eq.~(\ref{eq: diagonalization 1}) can be solved straightforwardly. The multiplier in front of the noise term \(F(z,\tau)\) in Eq.~(\ref{eq: one-dimensional field}) is given by:
\begin{equation}
\label{eq: one d noise multiplier}
     \xi(z,\tau)= \sqrt{n\left[\frac{\partial}{\partial \tau}+\Gamma\right]\big({\rho}_{uu}(z,\tau)-|{\rho}_{ul}(z,\tau)|^2\big)}.
\end{equation}
This expression differs from the corresponding one in the phenomenological approach outlined in Ref.~\cite{2000'Larroche}, which, in our notation, is:
\begin{equation*}
     \xi'(z,\tau)= \sqrt{n \Gamma{\rho}_{uu}(z,\tau)}.
\end{equation*}
{If the system is instantly excited, the stochastic polarization fields have the following nonzero initial conditions:
\begin{equation*}
    P_{\text{noise}}^{(+)}(z,0)=H(z)\sqrt{n\big({\rho}_{uu}(z,0)-|{\rho}_{ul}(z,0)|^2\big)}, 
\end{equation*}
where the Gaussian white noise term $H(z)$ has zero mean and satisfies the following correlation properties:
\begin{align*}
    \langle H^*(z)H(z')\rangle &= \frac{1}{S_\text{coh.}}\delta(z-z'),\\
    \langle H(z)H(z')\rangle &= 0.
\end{align*}
In the presented numerical study,} the atoms are initially in their ground states, {so} the {stochastic} polarization fields are initially zero, {$P_{\text{noise}}^{(+)}(z,0)=P_{\text{noise}}^{(-)}(z,0)=0$}.

The equations for the atomic degrees of freedom read as follows:
\begin{subequations}
\label{eq: 1d atoms}
\begin{align}
&\begin{multlined}
\label{eq: 1d atoms 1}
    \frac{\partial}{\partial \tau}\rho_{uu}(z,\tau) = ... \\+ i \left[\Omega^{(+)}(z,\tau) \rho_{lu}(z,\tau)-\Omega^{(-)}(z,\tau) \rho_{ul}(z,\tau)\right],
\end{multlined}\\
&\begin{multlined}
\label{eq: 1d atoms 2}
    \frac{\partial}{\partial \tau}\rho_{ll}(z,\tau) = ... \\- i \left[\Omega^{(+)}(z,\tau) \rho_{lu}(z,\tau)-\Omega^{(-)}(z,\tau) \rho_{ul}(z,\tau)\right],
\end{multlined}\\
&\begin{multlined}
\label{eq: 1d atoms 3}
    \frac{\partial}{\partial \tau}\rho_{ul}(z,\tau) = -\Gamma \rho_{ul}(z,\tau)/2 \\- i \Omega^{(+)}(z,\tau) \left[\rho_{uu}(z,\tau)- \rho_{ll}(z,\tau)\right],
\end{multlined}
\end{align}
\end{subequations}
where the contribution from rate equations (\ref{eq: rate equations for the pumping}) is denoted by dots. The remaining quantities are determined by complex conjugation:
\begin{equation*}
    \rho_{lu}(z,\tau)=\rho_{ul}^*(z,\tau), \quad \Omega^{(-)}(z,\tau)=\Omega^{(+)*}(z,\tau).
\end{equation*}

In the following, we study the superfluorescence in the three stages of spontaneous emission, amplified spontaneous emission (ASE), and saturation.

%%%%%%%%%%%%%%%%%%%%%%%%%%%%%%%%%%%%%%%%%%%%%%%%%%%%%%%%
\subsubsection*{Spontaneous emission}
%%%%%%%%%%%%%%%%%%%%%%%%%%%%%%%%%%%%%%%%%%%%%%%%%%%%%%%%%%%%%%%%
 \begin{figure*}[t!]

 % mention 3.4 in the caption
    \centering
    \includegraphics[width = \linewidth]{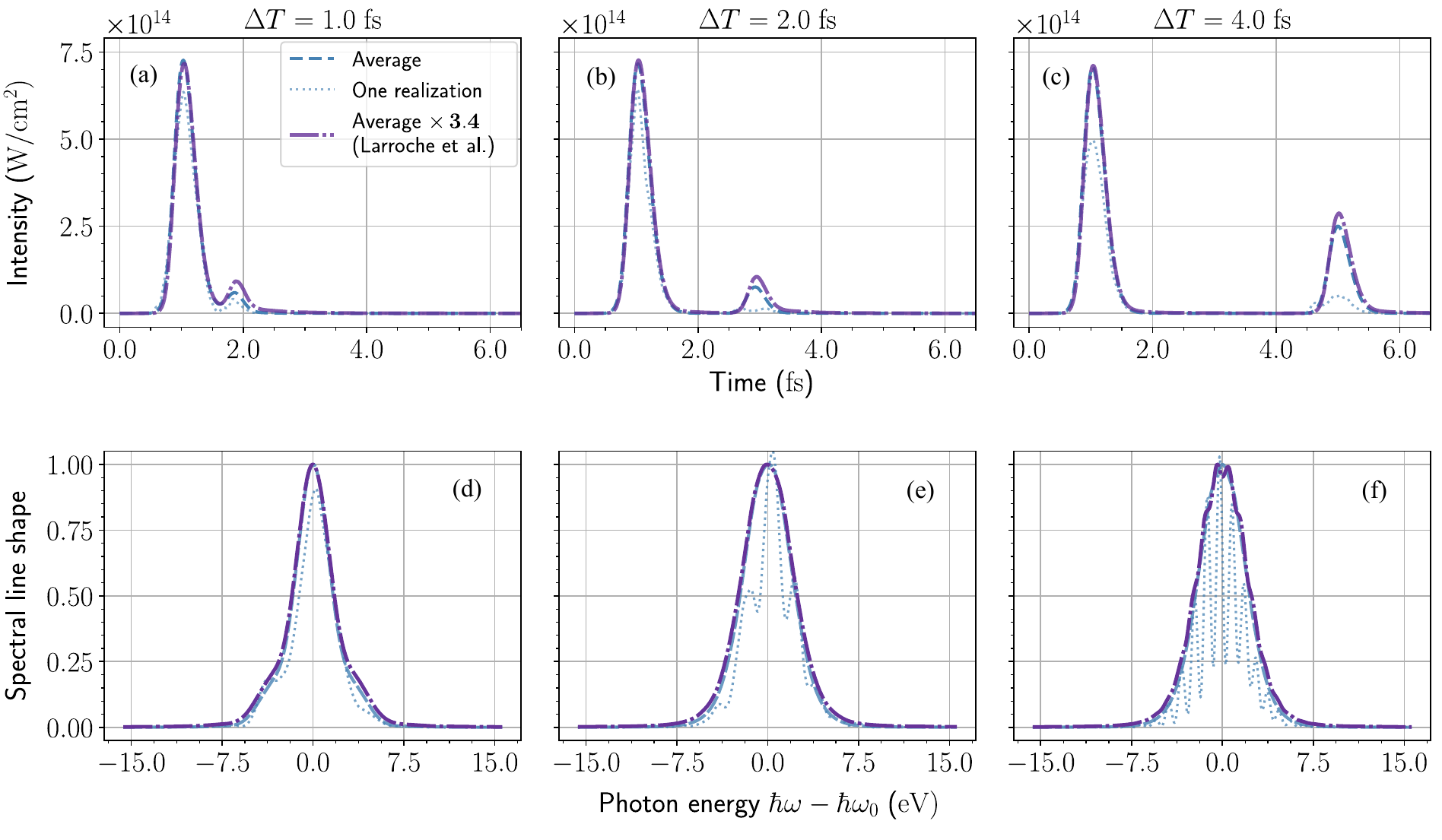}
    \caption{Temporal (a-c) and normalized spectral (d-f) intensities of amplified spontaneous emission at a length of $z=3.0L_g$ ($3.5$ \textmu m) for separations of $\Delta T = 1.0$ fs, $2.0$ fs, $4.0$ fs between the two pump pulses. The average temporal intensities based on our (dashed blue line) and phenomenological (dash-dotted purple line) methodologies almost coincide when scaled to the same peak intensity (the purple curves have been scaled by a factor of 3.4). Similarly, the normalized average spectral intensities based on our estimation (dashed blue line), and on the phenomenological methodology (dash-dotted purple line) are almost identical. The dotted blue lines show single realizations, which exhibit a fringe pattern in the emitted spectrum due to the temporally separated two emission bursts.}
    \label{fig: spectra and intensities for medium z}
\end{figure*}
%%%%%%%%%%%%%%%%%%%%%%%%%%%%%%%%%%%%%%%%%%%%%%%%%%%%%%%%%%%%%%%%

%%%%%%%%%%%%%%%%%%%%%%%%%%%%%%%%%%%%%%%%%%%%%%%%%%%%%%%%%%%%%%%%
 \begin{figure*}[t!]
    \includegraphics[width = \linewidth]{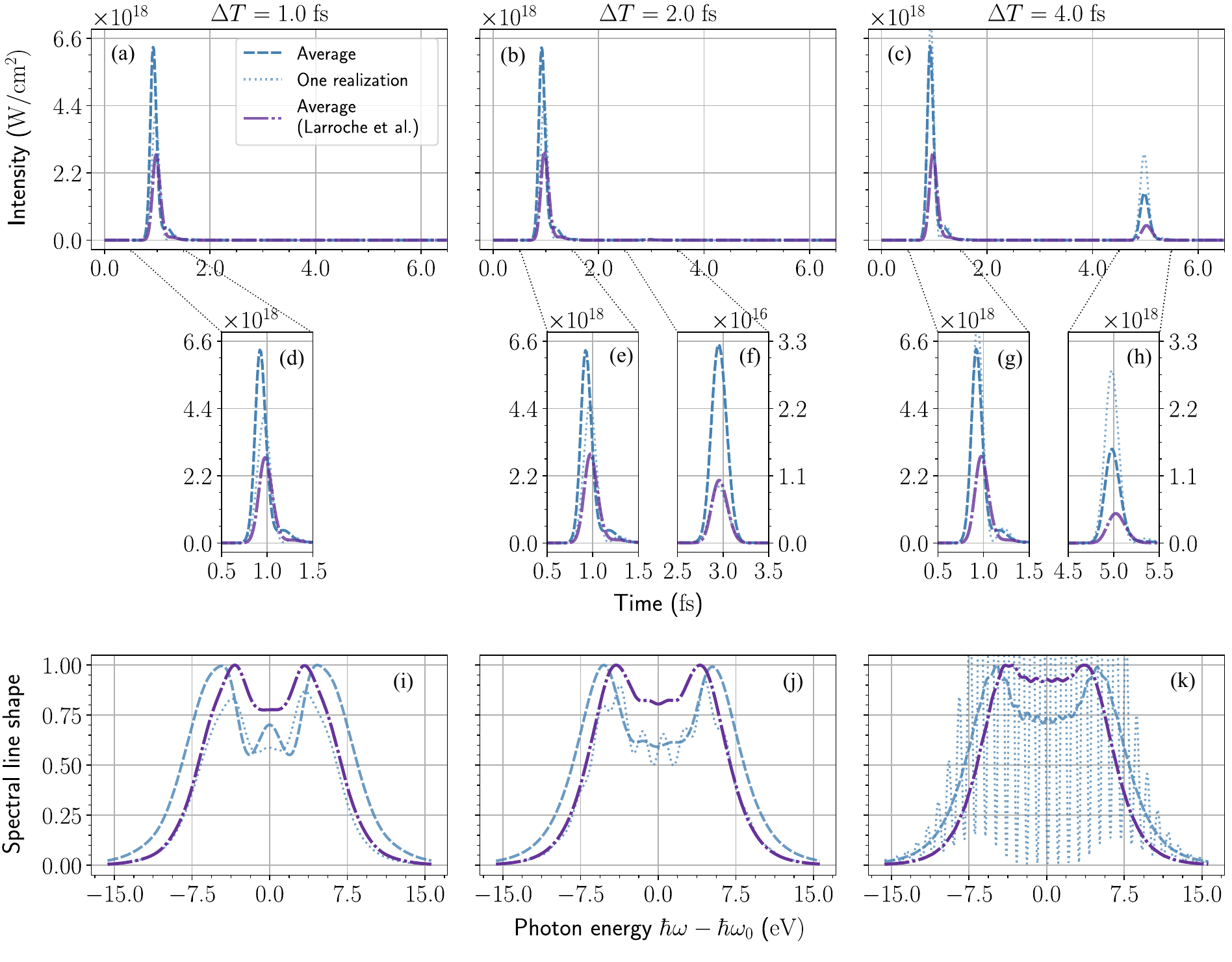}
    \caption{Temporal (a-c) and normalized spectral (i-k) intensities in the saturation regime at a length of $z=14.3L_g$ ($16.6$ \textmu m) for separations of $\Delta T = 1.0$ fs, $2.0$ fs, $4.0$ fs between the two pump pulses. The resulting two emission bursts differ drastically in their intensities; we show each burst in separate panels (d) to (h). The emission spectra feature the Autler-Townes splittings due to Rabi oscillations, and in the single-shot spectra, spectral fringes become apparent.}
    \label{fig: spectra and intensities for large z}
\end{figure*}
%%%%%%%%%%%%%%%%%%%%%%%%%%%%%%%%%%%%%%%%%%%%%%%%%%%%%%%%%%%%%%%%

%%%%%%%%%%%%%%%%%%%%%%%%%%%%%%%%%%%%%%%%%%%%%%%%%%%%%%%%%%%%%%%%
 \begin{figure*}[t!]

 % double check the concept of fluence
 
    \centering
    \includegraphics[width = \linewidth]{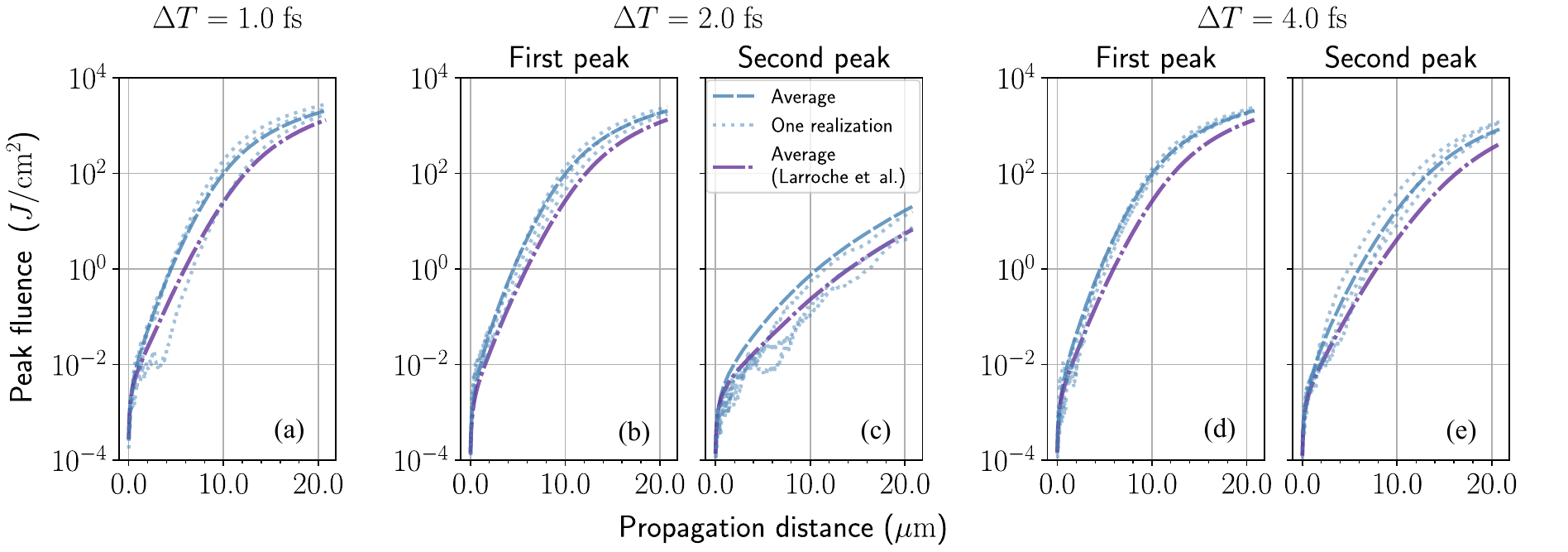}
    \caption{Emitted peak fluence (temporally integrated intensity) as a function of propagation distance through the gain medium (gain curves) for pump-pulse separations of $\Delta T = 1.0$ fs, $2.0$ fs, and $4.0$ fs. Each panel features gain curves based on single realizations (dotted blue lines) and averages of our methodology (dashed blue lines) and the phenomenological approach (dash-dotted purple line) for the first and (for $\Delta T =2.0$ fs and $4.0$ fs) second emission burst.}
    \label{fig: gain curves}
\end{figure*}
%%%%%%%%%%%%%%%%%%%%%%%%%%%%%%%%%%%%%%%%%%%%%%%%%%%%%%%%%%%%%%%%

The regime of spontaneous emission is illustrated in Fig. \ref{fig: spectra and intensities for first z}. Panels (a) to (c) demonstrate the time-dependent intensity of the emission at $z=35$ nm or $0.03L_g$, calculated based on Eq.~(\ref{eq: observables for the fields a}). Similarly, panels (d) to (f) depict the normalized spectra given by Eq.~(\ref{eq: observables for the fields b}). By construction, our formalism fully reproduces the expectation value of the intensity of spontaneous emission according to the following analytical expression:
\begin{equation}
\label{eq: exact sp}
    I_{\text{sp.}}(\Delta z,\tau) = \hbar\omega_0 \times\frac{3\Delta o}{8\pi}\Gamma_{\text{rad.}}\times \Delta zn\rho_{uu}(0,\tau).
\end{equation}

For comparison, we show the corresponding results obtained by phenomenological noise terms (dash-dotted purple lines). Despite the agreement in the spectral profiles, the temporal profiles show large differences from the exact solution and novel noise terms. In particular, the phenomenological noise terms do not reproduce the exact fast rise in intensity. The analytical expression of the intensity of spontaneous emission given by the phenomenological noise terms has the following form:
\begin{multline*}
    I_{\text{sp.}}'(\Delta z,\tau) = \hbar\omega_0 \times\frac{3\Delta o}{8\pi}\Gamma_{\text{rad.}}\\\times \Delta zn \int_0^\tau \frac{d\tau'}{\Gamma} \rho_{uu}(0,\tau') e^{-\Gamma(\tau-\tau')}.
\end{multline*}
The larger the coefficient $\Gamma$, the faster the intensity responds to changes in the upper-state population. In the opposite case, as $\Gamma \rightarrow 0$, the delay between population change and emission increases. Since our formalism provides a more sophisticated multiplier of the noise terms that also depends on the derivative, small values of $\Gamma$ do not lead to an incorrect temporal intensity profile.

Fig. \ref{fig: spectra and intensities for first z} also includes intensities and spectra based on single noise realizations. Notably, these can deviate significantly from their corresponding averages. Therefore, it is challenging to notice the difference between distinct models for the stochastic sources at the level of single shots. Noticeable qualitative differences can be observed only at the level of statistical averages.

%%%%%%%%%%%%%%%%%%%%%%%%%%%%%%%%%%%%%%%%%%%%%%%%%%%%%%%%
\subsubsection*{Amplified spontaneous emission (ASE)}
%%%%%%%%%%%%%%%%%%%%%%%%%%%%%%%%%%%%%%%%%%%%%%%%%%%%%%%%

In the subsequent phase, spontaneous photons propagate along the sample and are amplified by the inverted medium. Fig. \ref{fig: spectra and intensities for medium z} provides the temporal and spectral properties of the field at $z=3.5$ \textmu m or $3.0L_g$. The fields are not sufficiently strong to drive the populations in Eqs.~(\ref{eq: 1d atoms 1}) and (\ref{eq: 1d atoms 2}), but they are strong enough to affect the dynamics of the coherences in Eq.~(\ref{eq: 1d atoms 3}). The perturbed coherences induce positive feedback in the equations governing the field variables~(\ref{eq: field 1d}), resulting in self-amplification.

Fig. \ref{fig: spectra and intensities for medium z} reveals that the shape of the average temporal and spectral intensities remains consistent, irrespective of the chosen model for the noise terms. However, the phenomenological methodology underestimates the absolute value of the emitted intensity by a factor of roughly $3.4$. This discrepancy can be linked to the strong differences in the temporal emission profiles of the spontaneous emission, leading to reduced overlap between the population inversion and emission in the temporal domain for the phenomenological approach.

In contrast to the spontaneous emission regime, the single realizations of the intensities strongly follow the evolution of their respective averages, up to a small multiplicative factor. The spectra based on single realizations have an additional feature that has been experimentally observed \cite{2022'Zhang}: The atoms radiate two bursts that interfere and result in a fringe pattern (higher-frequency modulation) in single shot emission spectra. Originating from random spontaneous emission, the two emission bursts have random and statistically independent phases. Upon averaging over multiple stochastic realizations, these modulations disappear. The envelopes of the normalized single-shot spectra roughly follow their average.

Different columns in Fig.~\ref{fig: spectra and intensities for medium z} represent various time delays between the pump pulses ionizing the atoms. The dynamics triggered by the first pump pulse is identical in all three cases. The second pump pulse creates slightly different population inversions for different delays (see Fig. \ref{fig: populations}), which, however, produce noticeable differences in the emitted intensity profiles.

At this point, it becomes evident that the choice of stochastic source model significantly influences the temporal profiles of spontaneous emission. Consequently, this discrepancy results in an underestimation of the intensity of amplified spontaneous emission when employing phenomenological noise terms. However, it is noteworthy that the shape of spectral and temporal intensity profiles in the ASE regime exhibits a relatively minor dependence on the chosen stochastic source model. { In this regime, the choice of the model only impacts the total photon number.}

{As discussed in the Introduction, the original framework detailed in Sec. \ref{sec: stochastic formalism} would result in diverging trajectories during amplification. Specifically, all dynamic variables would reach infinity within a finite time span. To regularize these trajectories, the numerical simulations in Ref. \cite{benediktovitch2023stochastic} were conducted with a modified system of equations (details are given in \cite{benediktovitch2023stochastic}), lacking a rigorous derivation. We assume that the phenomenological regularization of the diverging trajectories employed in \cite{benediktovitch2023stochastic} produces qualitatively similar results to the current approach in the ASE regime. However, we anticipate that it results in inaccurate photon numbers.}

%%%%%%%%%%%%%%%%%%%%%%%%%%%%%
\subsubsection*{Saturation}
%%%%%%%%%%%%%%%%%%%%%%%%%%%%%

Upon further amplification, the field becomes intense enough to influence the populations in Eqs. (\ref{eq: 1d atoms 1}) and (\ref{eq: 1d atoms 2}), leading to a reduction in population inversion and inducing Rabi oscillations. This is manifested in saturated amplification. As shown in Fig. \ref{fig: spectra and intensities for large z}, the temporal and spectral properties become more complex for $z=16.6$ \textmu m or $14.3L_g$.

At first glance, the intensities and spectra calculated based on different stochastic sources behave differently. However, the difference is due to the slight delay in the amplification process caused by the phenomenological noise terms, as discussed previously. This is further confirmed by Fig.~\ref{fig: gain curves}, which shows the emitted peak fluence as a function of propagation distance (gain curve). In the exponential gain region, the gain curves of the two models differ by a constant shift. In the saturation regime, regardless of the utilized model of the noise terms, the temporal profiles of the fields are identical for propagation distances that are accordingly rescaled to match the peak fluence.

As depicted in Fig.~\ref{fig: populations}, varying delays between pump pulses result in different population inversions. In the case of the shortest delay of $\Delta T = 1.0$ fs, shown in Fig. \ref{fig: spectra and intensities for large z} (a), the two superfluorescent pulses merge into one intense peak. For a delay of $\Delta T = 2.0$ fs, we can observe the second peak centered around $3.0$ fs with almost the same width but two orders of magnitude less intense. Finally, for the delay of $\Delta T = 4.0$ fs, we can clearly see both peaks. The secondary peak is less intense but comparable in strength.

Similar to the ASE regime, the single-shot spectra illustrated in Figs. \ref{fig: spectra and intensities for large z} (j) and (k) exhibit high-frequency modulations attributed to the interference of two superfluorescent emission bursts.
% At the level of single realizations, we can clearly see the higher-frequency modulation of spectra shown in Fig. \ref{fig: spectra and intensities for large z} (e, f). This effect was also observed at the previous stage of the process (amplified spontaneous emission), see Fig. \ref{fig: spectra and intensities for medium z} (j, k).
% Fig.~\ref{fig: spectra and intensities for large z} (a) illustrates that with the shortest delay, the two superfluorescent pulses merge into one. As the delay increases, the superfluorescent pulses become well-separated, but their intensities differ significantly.
%%%%%%%%%%%%%%%%%%%%%%%%%%%%%%%%%%%%%%%%%%%%%%%%%%
\subsection{Amplified spontaneous vs stimulated emission}\label{Amplified spontaneous vs stimulated emission}
%%%%%%%%%%%%%%%%%%%%%%%%%%%%%%%%%%%%%%%%%%%%%%%%%%

%%%%%%%%%%%%%%%%%%%%%%%%%%%%%%%%%%%%%%%%%%%%%%%%%%%%%%%%%%%%%%%%
 \begin{figure*}[t!]
    \centering
    \includegraphics[width = \linewidth]{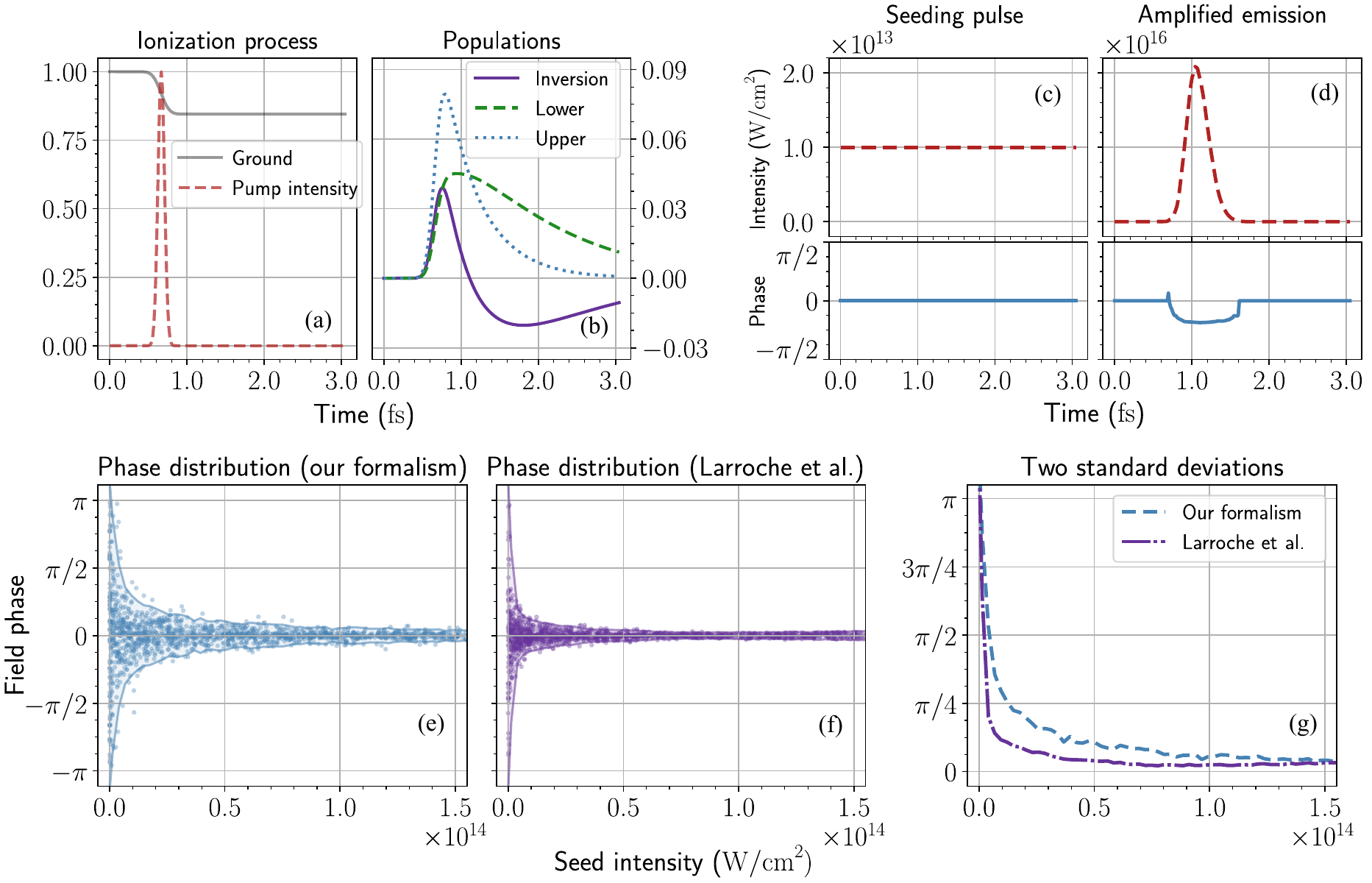}
    \caption{Study of the interplay between spontaneous and seeding emissions. Panels (a) and (b) illustrate the normalized temporal profile of the pump pulse along with the resulting dynamics of the atomic populations. Panels (c) and (d) provide an example of a seed pulse ($z=0.0$) and the outcome of its amplification ($z=6.0L_g$  or $7.0$ \textmu m). Panels (e) and (f) present the distributions of the phases for different intensities of the seeding emission. Panel (e) represents our formalism, whereas the simulations in panel (f) are based on the phenomenological formalism. Additionally, two standard deviations from both approaches are illustrated together in panel (g).}
    \label{fig: phases}
\end{figure*}
%%%%%%%%%%%%%%%%%%%%%%%%%%%%%%%%%%%%%%%%%%%%%%%%%%%%%%%%%%%%%%%%

The presented numerical study shows that the choice of noise terms primarily influences the expected number of emitted photons and does not heavily affect the qualitative properties of the superfluorescent emission. By introducing a narrow-band seed pulse (as it is possible in the case of the SACLA FEL \cite{Doyle:23, 2020'Kroll}), we can conduct a more detailed study of spontaneous emission. In particular, varying the seed intensity should directly impact the phase statistics of the amplified emission. A strong seed pulse will directly imprint its phase on the amplified fields. If the seed intensity is insufficient to dominate the spontaneous emission, the phase of the resulting emission will vary noticeably from shot to shot.

In this section, the pump pulse consists of a single short pulse. Figs. \ref{fig: phases} (a) and (b) illustrate the normalized temporal pulse profile along with the resulting atomic populations. The seed pulse possesses a stable phase and is long compared to the total upper-level decay time. The seed pulse is modeled as non-zero initial conditions for the field variables:
$$
    \Omega^{(+)}(0,\tau)=\Omega^{(-)*}(0,\tau)=\Omega_0,
$$
where $\Omega_0$ is strictly real and positive, setting a zero phase. Figs. \ref{fig: phases} (c) and (d) provide an example of a seed pulse and the outcome of its amplification.

Figs. \ref{fig: phases} (e) and (f) present the distributions of the phases of the emission burst as a function of different seed intensities, comparing our formalism (e) to the phenomenological noise terms (f). The two methodologies give noticeably different results. In particular, our method expects higher variations in the phases for small seed intensity and requires a stronger seed to suppress the effect of spontaneous emission, consistent with the underestimated intensity observed with phenomenological noise terms.

Notably, the narrow-band seed pulse and wide-band amplified field will interfere, producing fringes in single-shot spectra, similar to Figs.~\ref{fig: spectra and intensities for medium z} and \ref{fig: spectra and intensities for large z}. These fringes encode information about the relative phase of seed and emission burst. An experimental measurement of a large set of single spectra in this regime, and a comparison of the distribution function of the relative phases with our theoretical model, are envisioned and would reveal if single realizations of our theoretical approach indeed represent real quantum realizations.

%%%%%%%%%%%%%%%%%%%%%%%%%%%%%%%%%%%%%%%%%%%%%%%%%
\section{Discussion and conclusion}\label{finish}
%%%%%%%%%%%%%%%%%%%%%%%%%%%%%%%%%%%%%%%%%%%%%%%%%

In conclusion, we have proposed a comprehensive theoretical framework for describing x-ray superfluorescence by introducing novel noise terms to semiclassical Maxwell-Bloch equations. Our approach accurately reproduces the temporal and spectral properties of spontaneous emission. The method is relatively straightforward to implement in a numerical scheme, with minimal deviation from traditional Maxwell-Bloch equations or other existing phenomenological methods widely used for analyzing x-ray superfluorescence \cite{2000'Larroche, 2018'Krusic}. 

The numerical simulations show that the choice of noise terms primarily influences the expected number of photons and does not heavily affect the qualitative properties of the emission once it is amplified. However, it is important to mention that the examples were restricted to one-dimensional geometry. 

The ultimate benchmark of our framework would be comparison to experiments designed to measure observables that sensitively depend on the initial quantum fluctuations. Distributions of the total photon yield, pulse duration, and phase differences would be interesting to compare between theory and experiment since these distributions generally encode higher-order correlation functions of the emitted fields. Comparing the theoretical predictions to experimental data can offer insights into potential refinements of the current theory. 

The proposed formalism holds potential for further improvement and extension. Firstly, there is a possibility for improvement by considering noise terms beyond the first perturbation order. Secondly, the existing formalism imposes a constraint on the lifetimes of states, specifically requiring degenerate or nearly degenerate states to share exactly the same stationary lifetimes. {Additionally, the energy splitting between nearly degenerate states has been neglected.} However, any attempt to refine the current formulation might encounter the challenge of unstable solutions of the resulting equations. Despite the current limitations, our approach, based on ab initio derivations, opens the path to quantitative predictive modeling of x-ray superfluorescence.

\begin{acknowledgments}
We are grateful to Andrei Benediktovitch, Aliaksei Halavanau, and Uwe Bergmann for fruitful discussions. Special thanks to Andrei Benediktovitch for providing critical feedback on the manuscript. S.C. acknowledges the financial support of Grant-No.~HIDSS-0002 DASHH (Data Science in Hamburg-Helmholtz Graduate School for the Structure of the Matter). V.S. acknowledges the financial support of the Cluster of Excellence ``CUI: Advanced Imaging of Matter'' of the Deutsche Forschungsgemeinschaft (DFG) --- EXC 2056 --- project ID 390715994.
\end{acknowledgments}

\appendix

\section{Role of the quadratic noise terms}
\label{Role of the quadratic noise terms}

{
In Sec.~\ref{sec: stochastic formalism}, we introduced the Maxwell-Bloch equations (\ref{eq:atomic_equations}) and (\ref{eq: field equations}) augmented by noise terms. Notably, the stochastic contributions in Eq.~(\ref{eq:atomic_equations_4}) contain noise terms that linearly and quadratically depend on the atomic variables $\rho_{pq}(\textbf{r},\tau)$. These same linear and quadratic combinations of atomic variables can be found in the modified equations given in Sec.~\ref{sec: self}. Specifically, they enter Eq.~(\ref{eq: spontaneous correlator}), defining the statistical properties of the stochastic polarization fields $P_{s,\text{noise}}(\textbf{r},\tau)$ in Eq.~(\ref{eq: full P with noise}).

To illustrate the role of these linear and quadratic combinations introduced in Eq.~(\ref{eq:atomic_equations_4}) and involved in Eq.~(\ref{eq: spontaneous correlator}), we provide the following example. We consider a compact system of $N$ uncorrelated emitters prepared in identical states, namely $\rho_{pq}(\textbf{r},0)=\bar{\rho}_{pq}$. For simplicity, we suppose the volume of the system is much smaller than the wavelength of the emitted light, which results in simple expressions for the far-field amplitudes. Specifically, the corresponding Rabi frequencies $\Omega^{(\pm)}_s(\tau)$ become proportional to the integrals of the polarization fields, namely $\Omega^{(\pm)}_s(\tau) \sim \int P^{(\pm)}_s(\textbf{r},\tau) d\textbf{r}$.
Assuming the expressions for the polarization fields given in Eq.~(\ref{eq: full P with noise}), we write:
\begin{align*}
    \Omega^{(+)}_s(\tau) &\sim N \sum_{l,u} T_{lu,s} \bar{\rho}_{ul}+\int P^{(+)}_{s,\text{noise}}(\textbf{r},\tau) d\textbf{r},\\
    \Omega^{(-)}_s(\tau) &\sim N \sum_{l,u} T_{ul,s} \bar{\rho}_{lu}+\int P^{(-)}_{s,\text{noise}}(\textbf{r},\tau) d\textbf{r},
\end{align*}
where integration over the whole sample yielded the number of emitters $N$. In this section, we assume that the atomic variables $\rho_{pq}(\textbf{r},\tau)$ do not evolve in time and remain equal to $\bar{\rho}_{pq}$. In other words, we are interested in the properties of the fields emitted right after the preparation of the emitters. Based on Eq.~(\ref{eq: observables for the fields a}), we write the following expression for the emission intensity $I_s(\tau)$:}
\begin{widetext}

{
    
\begin{equation*}
    I_s(\tau) \sim \int \langle P^{(+)}_s(\textbf{r},\tau)P^{(-)}_s(\textbf{r}',\tau)\rangle d\textbf{r}d\textbf{r}'=N^2\sum_{l,u}  \bar{\rho}_{ul}T_{lu,s}\sum_{l',u'}  \bar{\rho}_{l'u'}T_{u'l',s}+\int \langle P^{(+)}_{s,\text{noise}}(\textbf{r},\tau)P^{(-)}_{s,\text{noise}}(\textbf{r}',\tau)\rangle d\textbf{r}d\textbf{r}'.
\end{equation*}
 The deterministic components of the polarization fields generate the term proportional to the squared number of the emitters $N^2$, as expected from the deterministic formalism. Assuming the statistical properties of the stochastic polarization fields given in Eq. (\ref{eq: spontaneous correlator}), we write: 
\begin{equation*}
\label{app: stochastic intensity}
    \int \langle P^{(+)}_{s,\text{noise}}(\textbf{r},\tau)P^{(-)}_{s,\text{noise}}(\textbf{r}',\tau)\rangle d\textbf{r}d\textbf{r}'=N\left[\,\sum_{\mathclap{u',u,l}}T_{lu,s} \bar{\rho}_{uu'}T_{u'l,s}\\-\sum_{l,u} \bar{\rho}_{ul}T_{lu,s}\sum_{l',u'} \bar{\rho}_{l'u'}T_{u'l',s}\right].
\end{equation*}
Here, we obtain the same linear and quadratic combinations of the atomic variables featured in Eq.~(\ref{eq:atomic_equations_4}) and involved in Eq.~(\ref{eq: spontaneous correlator}). The linear terms represent the intensity of spontaneous emission, whereas the quadratic terms provide the correction of the deterministic contribution. The total intensity of the field reads as follows:
\begin{equation}
\label{final eq appendix}
     I_s(\tau) \sim N(N-1)\sum_{l,u} \bar{\rho}_{ul}T_{lu,s}\sum_{l',u'} \bar{\rho}_{l'u'}T_{u'l',s}+N\sum_{\mathclap{u',u,l}}T_{lu,s}\bar{\rho}_{uu'}T_{u'l,s}.
\end{equation}
At high concentrations of the emitters, $N(N-1)$ is approximately $N^2$ so the correction provided by the quadratic noise terms can be safely neglected. 

Let us illustrate that the direct quantum-mechanical calculations produce the same results. Given the reduced density matrix of the emitters $\hat{\rho}(\tau)$, we express the light intensity as:
\[ I_s(\tau) \sim \text{Tr}\left[\hat{P}_s^{(-)} \hat{P}_s^{(+)}\hat{\rho}(\tau)\right]. \]
Here, we introduce the collective dipole-moment operators $\hat{P}_s^{(\pm)}$ as:
\[ \hat{P}_s^{(+)}=\sum_{a,u,l}T_{lu,s}\hat{\sigma}_{a,lu},\quad\quad
    \hat{P}_s^{(-)}=\sum_{a,u,l}T_{ul,s}\hat{\sigma}_{a,ul}, \]
where $\hat{\sigma}_{a,ul}=\ket{u}\bra{l}$ are transition operators between states $\ket{u}$ and $\ket{l}$ of emitter $a$. The expression for the light intensity then becomes:
\[ I_s(\tau) \sim \text{Tr}\left[\hat{P}_s^{(-)}\hat{P}_s^{(+)}\hat{\rho}(\tau)\right] = \sum_{l',u'} T_{u'l',s}\sum_{l,u} T_{lu,s}\sum_{a,b}\langle\hat{\sigma}_{a,u'l'}\hat{\sigma}_{b,lu}\hat{\rho}(\tau)\rangle, \]
where the terms in the sum over $a$ and $b$ can be categorized into two groups: $N(N-1)$ non-diagonal ($a\neq b$) terms and $N$ diagonal ($a=b$) terms:
\[ \sum_{a,b}\langle\hat{\sigma}_{a,u'l'}\hat{\sigma}_{b,lu}\hat{\rho}(\tau)\rangle=\sum_{a\neq b}\langle\hat{\sigma}_{a,u'l'}\hat{\sigma}_{b,lu}\hat{\rho}(\tau)\rangle+\sum_{a}\langle\hat{\sigma}_{a,u'l'}\hat{\sigma}_{a,lu}\hat{\rho}(\tau)\rangle=\sum_{a\neq b}\langle\hat{\sigma}_{a,u'l'}\hat{\sigma}_{b,lu}\hat{\rho}(\tau)\rangle+\delta_{l'l}\sum_{a}\langle\hat{\sigma}_{a,u'u}\hat{\rho}(\tau)\rangle. \]
Assuming that the emitters are prepared in identical states, $\langle\hat{\sigma}_{a,u'u}\rho(\tau)\rangle$ becomes $\bar{\rho}_{uu'}$. Additionally, assuming uncorrelated emitters, $\langle\hat{\sigma}_{a,u'l'}\hat{\sigma}_{a,lu}\rho(\tau)\rangle$ turns into the product $\bar{\rho}_{l'u'}\bar{\rho}_{ul}$. We maintain the assumption of stationary atomic variables. This yields the exact same expression for the light intensity as Eq. (\ref{final eq appendix}).}

\end{widetext}

% \bibliography{biblio}% Produces the bibliography via BibTeX.
\bibliography{bib}

%%%%%%%%%%%%%%%%%%%%%%%%%%%%%%%%%%%%%%%%%%%%%%%%%%%%%%%%%%%%%%%%
%%%%%%%%%%%%%%%%%%%%%%%%%%%%%%%%%%%%%%%%%%%%%%%%%%%%%%%%%%%%%%%%
%%%%%%%%%%%%%%%%%%%%%%%%%%%%%%%%%%%%%%%%%%%%%%%%%%%%%%%%%%%%%%%%

\end{document}